\UseRawInputEncoding 
\documentclass{aa} %

\usepackage[english]{babel}
\usepackage{soul}
\usepackage[utf8]{inputenc} 
\usepackage[section]{placeins}
\usepackage[T1]{fontenc} 
\usepackage{mathtools} 
\usepackage{float} 
\usepackage{graphicx} 
\usepackage[justification=centering]{caption} 
\usepackage{subcaption}
\usepackage{wallpaper}
\usepackage{nomencl}
\usepackage{fancyhdr}
\usepackage{url}
\usepackage{array}
\usepackage{xcolor}
\usepackage{rotfloat}
\usepackage{siunitx}
\usepackage{ulem}
\usepackage{enumitem}
\usepackage{lineno}

\usepackage{supertabular}
\usepackage{multirow}
\usepackage{tikz}
\usepackage{blindtext}
\usepackage{soul}
\usepackage{mathtools} 
\usepackage{float}
\usepackage{graphicx}
\usepackage{multicol}
\usepackage{amssymb}
\usepackage{comment}

\usepackage{txfonts}
\usepackage{xcolor}
\def\cmk#1{\textcolor{magenta}{(\bf matthieu: \bf#1)}}
\begin{document}

   \title{
   Quantifying the diffusion of suprathermal electrons by whistler waves between 0.2 and 1 AU with Solar Orbiter and Parker Solar Probe
   }


 \titlerunning{ Diffusion of suprathermal electrons by whistler waves between 0.2 and 1 AU}


   \author{L. Colomban\inst{1}, M. Kretzschmar\inst{1}, V. Krasnoselkikh\inst{1,2},  O. V. Agapitov \inst{2,3}, C. Froment\inst{1}, M. Maksimovic \inst{4}, M. Berthomier \inst{5}, Yu. V. Khotyainsev \inst{6}, D. B. Graham \inst{6} and S. Bale \inst{2,7}}

   \institute{LPC2E, CNRS/University of Orléans/CNES, 3A avenue de la Recherche Scientifique, Orléans, France\\
              \email{lucas.colomban@cnrs-orleans.fr}
         \and
             Space Sciences Laboratory, University of California, Berkeley, CA, USA
         \and
              Astronomy and Space Physics Department, National Taras Shevchenko University of Kyiv, Kyiv, Ukraine
        \and
             LESIA, Observatoire de Paris, Université PSL, CNRS, Sorbonne Université, Université de Paris, Meudon, France
          \and 
            LPP, CNRS, Ecole Polytechnique, Sorbonne Université, Observatoire de Paris, Université Paris-Saclay, Palaiseau, Paris, France        
          \and 
           Swedish Institute of Space Physics (IRF), Uppsala, Sweden
         \and 
             Physics Department, University of California, Berkeley, CA, USA
             }

\date{Received / Accepted}

 
\abstract
   {The evolution of the solar wind electron distribution function with heliocentric distance
   exhibits different features that are still unexplained, in particular, the fast decrease of the electron heat flux and the increase of the Strahl pitch angle width. Wave-particle interactions between electrons and whistler waves are often proposed to explain these phenomena. }
 {We aim at quantifying the effect of whistler waves on suprathermal electrons as a function of heliocentric distance.}
  {We first perform a statistical analysis of whistler waves (occurrence and properties) observed by Solar Orbiter and Parker Solar Probe between 0.2 and 1 AU. The wave characteristics are then used to compute the diffusion coefficients for solar wind suprathermal electrons in the framework of quasi-linear theory. These coefficients are integrated in order to deduce the overall effect of whistler waves on electrons along their propagation.}
   {About $110,000$ whistler wave packets are detected and characterized in the plasma frame, including their direction of propagation with respect to the background magnetic field and their radial direction of propagation. Most waves are aligned with the magnetic field and only $\sim 0.5 \%$ of them have a propagation angle greater than $45^\circ$. Beyond 0.3 AU, almost exclusively quasi-parallel waves propagating anti-sunward (some of them are found sunward but are within switchbacks with a change of sign of the radial component of the background magnetic) are observed. These waves are therefore Strahl-aligned and not counter-streaming. At 0.2 AU we find both Strahl-aligned and counter-streaming quasi-parallel whistler waves.}
 {Beyond 0.3 AU, the integrated diffusion coefficients show that the observed waves can explain the measured Strahl pitch angle evolution and are effective in isotropizing the halo. Strahl diffusion is mainly due to whistler waves with an angle of propagation $\theta \in [15,45]^\circ$ (the origin of which has not yet been fully determined). Near 0.2 AU, counter-streaming whistler waves can diffuse the Strahl electrons more efficiently than the Strahl-aligned waves by two orders of magnitude. }


   \authorrunning{Colomban et al.}
   
      \maketitle


\section{Introduction}

In the solar wind, it is common to distinguish three categories of electron populations. The core represents the thermal electrons at low energies and makes up around 95\% of the electrons in the solar wind \citep{pilipp_characteristics_1987,maksimovic_radial_2005,stverak_radial_2009}. The core electrons are affected by Coulomb collisions and thus have close to Maxwellian distributions. The suprathermal electrons of the halo have higher energies \citep{feldman_solar_1975,feldman_characteristic_1978,lazar_characteristics_2020} and are often represented by kappa distributions \citep{scudder_causes_1992,scudder_why_1992,maksimovic_kinetic_1997,pierrard_kappa_2010,lazar_towards_2015,pierrard_implications_2022}. The Strahl is a beam of high-energy electrons that follows the magnetic field lines, propagating in the anti-sunward direction \citep{rosenbauer_survey_1977,rosenbauer_preliminary_1976,pilipp_characteristics_1987,hammond_variation_1996}. 


The relative proportions of these populations evolve with radial distance. Indeed, the fraction density of halo electrons increases with distance while the fraction density of Strahl electrons decreases \citep{maksimovic_radial_2005,stverak_radial_2009}, suggesting a transfer from the Strahl to the halo. The Strahl population even disappears completely beyond 5.5 AU \citep{graham_evolution_2017}. Moreover, the Strahl Pitch Angle Width (PAW) increases with heliocentric distance \citep{hammond_variation_1996,graham_evolution_2017,bercic_scattering_2019}. Such changes are unexpected when considering an adiabatic propagation, that is to say, that conserves the first adiabatic invariant and the energy. Indeed, an adiabatic propagation predicts the focusing of the electrons along the magnetic field lines, and a Strahl strengthening \citep{lemaire_model_1970,maksimovic_kinetic_1997,pierrard_electron_1999,bercic_scattering_2019}. Because of the strong decrease of the collision cross-section with electron energy, collisions cannot explain the observed behavior for energies above 250 eV \citep{boldyrev_kinetic_2019,bercic_interplay_2021}. \\
\cite{macneil_radial_2020} showed that the Strahl's PAW is larger when the Strahl is directed sunward, which is the case during switchbacks. As electrons that encounter switchbacks travel larger distances, this result suggests that there is a nearly constant mechanism that diffuses these electrons in pitch angle. 
Moreover, \cite{bercic_scattering_2019} revealed a correlation between the PAW and the plasma $\beta_{\rm ec||}$ (ratio between the parallel thermal pressure of the core electrons and the magnetic pressure), which is an indication that a collisionless process can explain the Strahl's broadening. 
Wave-particle interactions have been suggested as a possible mechanism to broaden the Strahl and to diffuse the electrons from the Strahl into the halo \citep{feldman_characteristic_1978,pilipp_characteristics_1987}. 

Another unsolved problem is the decrease of the solar wind heat flux with heliocentric distance. The heat flux is the third moment of the distribution function and is mainly carried by electrons because they are the lightest particles in the solar wind \citep{feldman_solar_1975,marsch_kinetic_2006}. More precisely, it is carried by the two suprathermal components of the electron populations, the Strahl and the halo, because of their high energies and velocity drifts in the proton reference frame. Indeed, in this frame, the core has a sunward bulk velocity while the halo and the Strahl have an anti-sunward bulk velocity \citep{feldman_solar_1975,scime_regulation_1994}. These drift speeds satisfy the zero current condition. When a Strahl beam is clearly present in the observed distribution function it has the dominant contribution to the heat flux \citep{pilipp_characteristics_1987}. The Strahl is often present close to the Sun \citep{halekas_electrons_2020,halekas_electron_2021} and in a fast solar wind \citep{fitzenreiter_observations_1998,stverak_radial_2009}. On the other hand, around 1 AU and beyond, the Strahl is not always observed, especially in the slow solar wind \citep{graham_evolution_2017,gurgiolo_absence_2017}.\\
The solar wind heat flux decreases with the heliocentric distance \citep{feldman_characteristic_1978,scime_regulation_1994,stverak_electron_2015,halekas_electron_2021,pierrard_implications_2022}. 
 A free expansion would result in the electron heat flux decreasing following the magnetic field amplitude, that is as r $^{-2}$ near the Sun and as $r^{-1}$ far from the Sun \citep{scime_regulation_1994}. However, the observed exponent is rather between $-3$ and $-2.4$ \citep{scime_regulation_1994,stverak_electron_2015,halekas_electrons_2020}. It is therefore necessary to introduce additional mechanisms for heat flux dissipation, which can be collisions or wave-particle interactions. 
 
 A fully collisional closure relationship for the electron heat flux is given by
   \begin{equation}
 q_{\rm e}=-\kappa  \nabla T_{\rm e}, 
 \label{heat_flux}
   \end{equation}
 where $q_{\rm e}$ is the electron heat flux in $\si{\watt}/\si{\meter^2}$, $T_{\rm e}$ is the electron temperature ($\si{\kelvin})$ and $\kappa$ is the electron thermal conductivity. However, this relation does not work in the solar wind, as
 \cite{scime_regulation_1994} showed that it predicts a decay in $r^{-4.6}$, which is too fast, and electron heat flux values significantly larger than the ones observed for $r<1$ $AU$. 
 The Knudsen number is the ratio between the mean free path and the scale of the temperature gradient. \cite{bale_electron_2013} showed that for Knudsen numbers larger than 0.28 the electron heat flux (divided by the saturation heat flux) is no longer proportional to the Knudsen number as predicted by Equation \ref{heat_flux}.  This was recently confirmed by \cite{halekas_electron_2021} who, using a more realistic temperature law and Parker Solar Probe observations, showed that the observed heat flux is always smaller than the one predicted by Equation \ref{heat_flux}. It is therefore clear that a purely collisional mechanism cannot explain alone the decrease of the heat flux. \cite{halekas_electron_2021} also showed that the dissipation of the electron heat flux is correlated with the plasma $\beta$, which is here again an indication that a collisionless mechanism is regulating the heat flux. Therefore, wave-particle interactions are most likely necessary to explain both the radial evolution of the electron populations and the decrease of the heat flux. \\

 Electron-scale waves such as whistler waves are commonly observed in the solar wind below 1 AU \citep{gurnett_plasma_1977,neubauer_fast_1977,lacombe_whistler_2014,tong_statistical_2019,jagarlamudi_whistler_2020,kretzschmar_whistler_2021,cattell_parker_2021,froment_whistler_2023} and are natural candidates for explaining the radial evolution of the electron velocity distribution. Whistler waves are electromagnetic, mostly right-hand circularly-polarized waves with a frequency in the plasma frame that is between the lower hybrid frequency ($f_{\rm LH}$) and the electron cyclotron frequency ($f_{\rm ce}$). One can distinguish three categories of whistler waves, depending on their direction of propagation with respect to both the background magnetic field and the radial direction: quasi-parallel anti-sunward whistler waves, anti-sunward oblique whistler waves, and sunward whistler waves. These characteristics, together with the wave amplitude, are required to evaluate the efficiency of the waves to diffuse electrons. Note that we use the terms parallel or aligned to designate both a parallel or anti-parallel propagation with respect to the background magnetic field. We now briefly review the observations and theoretical results for the different types of whistler waves. \\

Small amplitude, quasi-parallel anti-sunward whistler waves are commonly observed in the solar wind \citep{lacombe_whistler_2014,stansby_experimental_2016,kajdic_suprathermal_2016,tong_statistical_2019,chust_observations_2021,kretzschmar_whistler_2021}. Theoretical studies \citep{gary_electron_1975,gary_solar_1977,shaaban_clarifying_2018,shaaban_quasi-linear_2019} and simulations \citep{lopez_particle--cell_2019,micera_particle--cell_2020,micera_role_2021} have shown that they can be generated by the heat flux instability (WHFI). This has been confirmed by \cite{tong_whistler_2019} with in situ observations. In this case, whistler waves are produced because of the heat flux that results from the drift between the core and halo electrons. The waves resonate with the halo electrons (first normal resonance) and diffuse them in pitch angle. 
The temperature anisotropy instability (TAI) can also generate small amplitude quasi-parallel whistler waves \citep{sagdeev_instability_1960,kennel_limit_1966,gary_whistler_1996,stverak_electron_2008,saito_all_2007,lazar_instability_2011,lazar_electromagnetic_2013,lazar_interplay_2014,lazar_towards_2015,lazar_electromagnetic_2018,lazar_whistler_2019,jagarlamudi_whistler_2020,vasko_quasi-parallel_2020}. The drift between the different populations favor the generation of anti-sunward waves at 1 AU \citep{vasko_quasi-parallel_2020}. \cite{sarfraz_combined_2020} discussed the combined role of these two instabilities for the generation of quasi-parallel anti-sunward whistler waves. Recently, \cite{bercic_whistler_2021} proposed a third generation mechanism based on the suprathermal electron deficit in the anti-Strahl direction \citep{pilipp_characteristics_1987,halekas_electrons_2020,halekas_electron_2021,halekas_sunward_2021,halekas_radial_2022,bercic_coronal_2020}. In that case, electrons diffuse to fill the suprathermal deficit.\\
Quasi-parallel anti-sunward whistler waves produced by the WHFI have been proposed to regulate the heat flux by diffusing the halo electrons, which is supported by both theoretical analyses and observations \citep{gary_electron_1975,feldman_evidence_1976,gary_solar_1977,scime_regulation_1994,gary_electron_1999,lacombe_whistler_2014}.
However, \cite{tong_statistical_2019}, based on Artemis observations, questioned the role of this type of whistler waves because of their small amplitude and suggested that additional work was needed. 
 Similarly, \cite{kuzichev_nonlinear_2019} used a particle-in-cell code simulations and showed that whistler waves do not significantly suppress the electron heat flux. \\
 \cite{shaaban_quasi-linear_2019} used quasi-linear theory to conclude that the heat flux instability only slightly modifies the drift between the different components. Therefore, the role of quasi-parallel anti-sunward whistler waves in reducing the heat flux is not clear. Furthermore, since these waves interact mainly with the halo, their role in diffusing the Strahl has been little studied. \citet{pierrard_evolution_2011} have adopted a kinetic approach by solving the Fokker-Planck equation and adding a term taking into account a turbulent whistler wave spectrum. Quasi-linear scattering turbulence is due to interaction with aligned whistlers, whose intensity verifies $(B{\rm w}/B_{\rm 0})^2=0.01$. This study indicates that assuming this turbulence spectrum, whistlers can determine halo formation. \\

Anti-sunward oblique whistler waves have been proposed to regulate the heat flux, especially close to the Sun \citep{pistinner_self-inhibiting_1998,komarov_self-inhibiting_2018,roberg-clark_suppression_2018,micera_particle--cell_2020,micera_role_2021,halekas_electron_2021,cattell_parker_2021,cattell_narrowband_2021}. Oblique waves ($\sim 70^\circ$) can be generated by the fan instability \citep{kadomtsev_electric_1968,parail_kinetic_1978,roberg-clark_scattering_2019}, which, in the solar wind, is driven by the Strahl electrons (first anomalous resonance) \citep{krafft_interaction_2003,vasko_whistler_2019,verscharen_self-induced_2019}. Theoretical analysis and simulations have shown that the fan instability can diffuse the Strahl \citep{verscharen_self-induced_2019,micera_particle--cell_2020,micera_role_2021,cattell_modeling_2021}. Indeed, oblique whistler waves have a larger left-hand polarized and electrostatic components and can interact with the Strahl electrons more effectively. Oblique waves ($\sim 70^\circ$) have been observed at 1 AU by \cite{cattell_narrowband_2020} using STEREO data in association with stream interaction regions, coronal mass ejections, and interplanetary shocks. Using cross spectra of Parker Solar Probe's first encounter ($\sim$ 0.2 AU), \cite{froment_whistler_2023} showed that $3\%$ of observed waves had an oblique propagation angle ($\ge 45^\circ$). To our knowledge, there are very few observations of oblique whistler waves in the free solar wind beyond 0.3 AU. This is in line with the results of \cite{jeong_stability_2022}. Indeed, in this study, they used fan-like instability thresholds derived by \cite{verscharen_self-induced_2019} and Strahl properties measured by Parker Solar Probe and Helios to show that Strahl electrons are on average stable against fan instability between 0.1 and 1 AU. These results suggest that this instability is probably very rare in the solar wind (in this heliocentric distance range) and that, if it exists, it can only be excited sporadically. This also agrees with the results of the kinetic stability analysis carried out by \cite{schroeder_stability_2021}.  \\

Sunward whistlers are particularly interesting because they can interact very efficiently with the Strahl \citep{vocks_electron_2005,saito_all_2007,sarfraz_combined_2020,cattell_modeling_2021}. \cite{saito_all_2007} proposed a wave/wave interaction mechanism to produce quasi-aligned sunward whistler waves. \cite{vasko_quasi-parallel_2020}, using typical properties of the electron distribution function at 1 AU, showed that quasi-parallel sunward whistler waves can be generated by the temperature anisotropy instability and are expected to have smaller frequencies, wave numbers, and growth rates than the anti-sunward ones. 
An interval containing sunward and anti-sunward whistler waves was observed in association with a magnetic flux rope by \citep{lacombe_whistler_2014} using CLUSTER data. In this study, they proposed that wave generation in these two directions is due to the presence of a bi-directional electron distribution associated with the flux rope. However, the absence of electron distribution function measurements during this interval makes it impossible to verify this hypothesis.
A case of sunward propagating whistler wave packet has been studied by \cite{agapitov_sunward-propagating_2020} in association with a magnetic field dip at a switchback \citep{bale_highly_2019,krasnoselskikh_localized_2020,dudok_de_wit_switchbacks_2020,agapitov_flux_2022} boundary using data from Parker Solar Probe (PSP) Encounter 1 at $\sim 0.2$ AU ($\simeq$ 43 $R_{\odot}$). This wave packet contained subpackets with propagation angles varying from quasi-parallel to oblique, probably due to propagation in an inhomogeneous background magnetic field.
\cite{froment_whistler_2023} also studied some examples of sunward waves during Encounter 1 of PSP and found that whistler waves (without making the distinction sunward /anti sunward) are associated with magnetic field dips in $64\%$ of the cases. Using burst waveforms observed during Encounter 1 of PSP, \cite{karbashewski_whistler_2023} reported both sunward, anti-sunward, and counter-propagating (propagating in both directions) whistlers. The waves observed were predominantly quasi-parallel, confirming the results of \cite{cattell_narrowband_2020,froment_whistler_2023}. 
\cite{karbashewski_whistler_2023} suggested that the generation of these waves is related to the temperature anisotropy. A shift in the distribution functions of the electrons trapped in the dips, caused by the propagation of these structures, would favor the generation of sunward waves. These sunward waves are theoretically expected to have higher frequencies than the anti-sunward waves. It's important to note that the distribution functions considered by \cite{karbashewski_whistler_2023} are different from the typical 1 AU distribution functions used by \cite{vasko_quasi-parallel_2020}. In particular, the supposed drift of trapped electrons due to magnetic dip motion explains the difference in predicted frequencies for sunward versus anti-sunward waves in these two studies. \\



Several observational studies have attempted to determine where and with what properties whistler waves occur in the heliosphere.
An anti-correlation between the solar wind speed (above 300 $\si{\kilo \meter}/\si{\second}$) and the occurrence of whistler waves has been found by \cite{lacombe_whistler_2014} and \cite{jagarlamudi_whistler_2020,jagarlamudi_whistler_2021}. \cite{jagarlamudi_whistler_2020} found a decrease in the whistler waves occurrence when getting closer to the Sun (from 1 AU to  0.3 AU) while \cite{kretzschmar_whistler_2021} found an increase of the whistler occurrence from 1 AU to 0.5 AU. \cite{jagarlamudi_whistler_2021,cattell_parker_2021} used electromagnetic and particle data of Parker Solar Probe to show evidence of diffusion of the Strahl electrons by whistler waves. In addition, \cite{cattell_parker_2022} has shown, using Parker Solar Probe data from Encounters 1 through 9, that whistler waves are rarely observed inside $\sim$ 0.13 AU ($\simeq$ 28 $R_{\odot}$, Solar radii). Furthermore, \cite{kajdic_suprathermal_2016} used CLUSTER data to show that the presence of whistlers (which were in large majority quasi-aligned) was correlated to a higher Strahl PAW. The observation of increased Strahl PAW during broadband whistler fluctuations was also observed by \cite{pagel_scattering_2007} using ACE data. \\

Nevertheless, despite these various observations suggesting the important role of whistler waves in solar wind electron diffusion, a quantification of the effect of whistler waves on suprathermal electrons in the solar wind is still missing.



In this study we aim to answer whether whistler waves can explain the transfer of electrons from the Strahl to the halo, and if yes, what are the characteristics of the waves responsible for it.
To do so, we first analyze Solar Orbiter and Parker Solar Probe data to determine the whistler wave properties between 0.2 and 1 AU. Then, we use the quasi-linear theory to compute the efficiency of these waves in diffusing the electron populations. Section \ref{Whistler waves statistics} presents the identification of the whistler waves and the determination of their properties. In Section \ref{Diffusion of solar wind electrons by Whistler waves}, we compute both the local and integrated effect of the whistler waves on the electron diffusion. In Section \ref{discussion}, we discuss our results.

\section{Whistler waves statistics}
\label{Whistler waves statistics}
\subsection{Data and Analysis} 
\label{Data and Analysis}
\subsubsection{Solar Orbiter and Parker Solar Probe Data}
We use data from the Solar Orbiter mission \citep{mueller_solar_2013,muller_solar_2020} obtained between July 2020 and March 2022, and covering distances between 0.3 AU and 1 AU (four perihelia). We exclude the two Venus flybys. The waves are identified and characterized using the Radio and Plasma Waves (RPW) experiment \citep{maksimovic_solar_2020}. RPW uses a Search Coil Magnetometer \citep{jannet_measurement_2021} and three electric antennas to produce both snapshot and continuous waveforms (SWF and CWF) of the fluctuating magnetic and electric fields (we use waveforms at 256 $\si{\hertz}$). Both the consideration of the CWF data products in addition to the SWF and the extension of the considered period is a significant improvement over the analysis of \cite{kretzschmar_whistler_2021}. We also utilize data from the magnetometer (MAG) \citep{horbury_solar_2020} and the Solar Wind analyzer (SWA) \citep{owen_solar_2020} instruments to retrieve the background magnetic field ($\Vec{B_{\rm 0}}$ in the following) and the proton moments (density and velocity) of the solar wind. When available, we use the electron density delivered by RPW as a level 3 data product \citep{khotyaintsev_density_2021}. We also use 3D normal mode velocity distribution functions of EAS 1 and EAS 2 \citep{owen_solar_2020}. 

For Parker Solar Probe \citep{fox_solar_2016,raouafi_parker_2023}, the whistler waves are characterized using the FIELDS instrument that measures the DC and AC electric and magnetic fields \citep{bale_fields_2016,malaspina_digital_2016}. Specifically, we use continuous waveforms (acquired at 297.97 $\si{\hertz}$) and burst waveforms (150 000 $\si{\hertz}$) data products from November 1st to 11th, 2018. We limit ourselves to the first perihelion of Parker Solar Probe because of the change in the response in one of the SCM components for the following encounters, making it impossible to compute directly the polarization properties of the waves \citep{dudok_de_wit_first_2022}. A method using the two available components of the magnetic and electric fields has recently been proposed by \cite{colomban_reconstruction_2023} to recover whistler polarization properties despite this technical issue. This method will eventually allow us to extend the wave statistics to the other encounters of Parker Solar Probe. We also use the SWEAP Solar Probe Cup (SPAN-C) L3 data \citep{kasper_solar_2016,case_solar_2020} for the density and solar wind speed. In the following, we round the heliocentric distance of Parker Solar Probe during the first encounter ($\in$ [0.16,0.25] AU) to 0.2 AU. \\

 \subsubsection{Detection and characterization}
 \label{Detection and characterisation}
 Whistler waves are detected in the same way as in \cite{kretzschmar_whistler_2021}. 
 In short, this consists of locating the periods of time where the magnetic field fluctuations are at least twice the median fluctuations measured in data from the same day over at least 2 $\si{\hertz}$. If such a bump is detected, we define a wave packet as the part of the band-pass filtered time series with fluctuations larger than the upper quartile. A wave packet must be composed of at least four periods. We then check that the coherence, planarity, and polarization of the wave are all greater (using the spectral energy content weighted average) than 0.6 over the considered frequency range \citep{santolik_singular_2003}. For continuous waveforms, if there are more than 8h of data, we split the day into four and four different median spectra are used. This allows us to take into account possible variations in turbulence levels. Although we note that the level of turbulence is in the overwhelming majority of cases well below the measured amplitude of the waves.
The propagation angle with respect to the background magnetic field ($\theta$ in the following) is determined using both a minimum variance analysis \citep{sonnerup_magnetopause_1967,sonnerup_minimum_1998} on the filtered magnetic field waveforms and the analysis of the computed magnetic spectral matrices \citep{means_use_1972,santolik_singular_2003,taubenschuss_wave_2019}. These methods give the propagation angle with an ambiguity of $\pm 180 ^\circ$, which is next removed by determining the radial direction of propagation using the electric field. Unless otherwise specified we work in spacecraft reference frames; for Solar Orbiter, $\vec{X}$ points to the sun, $\vec{Z}$ points to the north, and $\vec{Y}$ completes the direct reference frame \citep{maksimovic_solar_2020}. The direction of propagation (sunward or anti-sunward) indicates whether the wave is mainly directed toward the Sun or not. The direction of propagation is determined by computing the sign of the X component of the Poynting vector $\vec{S}$ (Equation \ref{Poyting}). We make the assumption of the weak phase deviation proposed in \cite{kretzschmar_whistler_2021} to take into account the nearly constant instrumental phase shift between the electric and magnetic field of Solar Orbiter. This assumption, as well as the precautions taken for the determination of the propagation direction, are detailed in appendix \ref{Annexe A}.  \\

We finally check for the right or left handed polarization (in the spacecraft frame) in the following way. For a right handed polarization, describing the magnetic field fluctuations $\vec{B}_{\rm w}(t)$ in a direct orthogonal Cartesian reference frame where $\vec{Z}$ is along the wave vector $\vec{k}$, we expect a phase shift of $\pm$ 90 $\si{\degree}$ between the X and Y components (if $\vec{k}$ and $\vec{B_{0}}$ are aligned and anti-aligned, respectively). 
This is the case, for anti-sunward waves and for sunward waves with $V_{\varphi} \geq V_{\rm SW_k}$ (where $V_{\varphi}$ is the phase velovity in the plasma frame and $V_{\rm SW_k}$ the solar wind speed along the $\vec{k}$ direction). Sunward whistler waves propagating with $V_{\varphi} < V_{\rm SW_k}$ have a left handed polarization in the spacecraft frame.

\begin{figure}
   \centering
   \includegraphics[width=\linewidth]{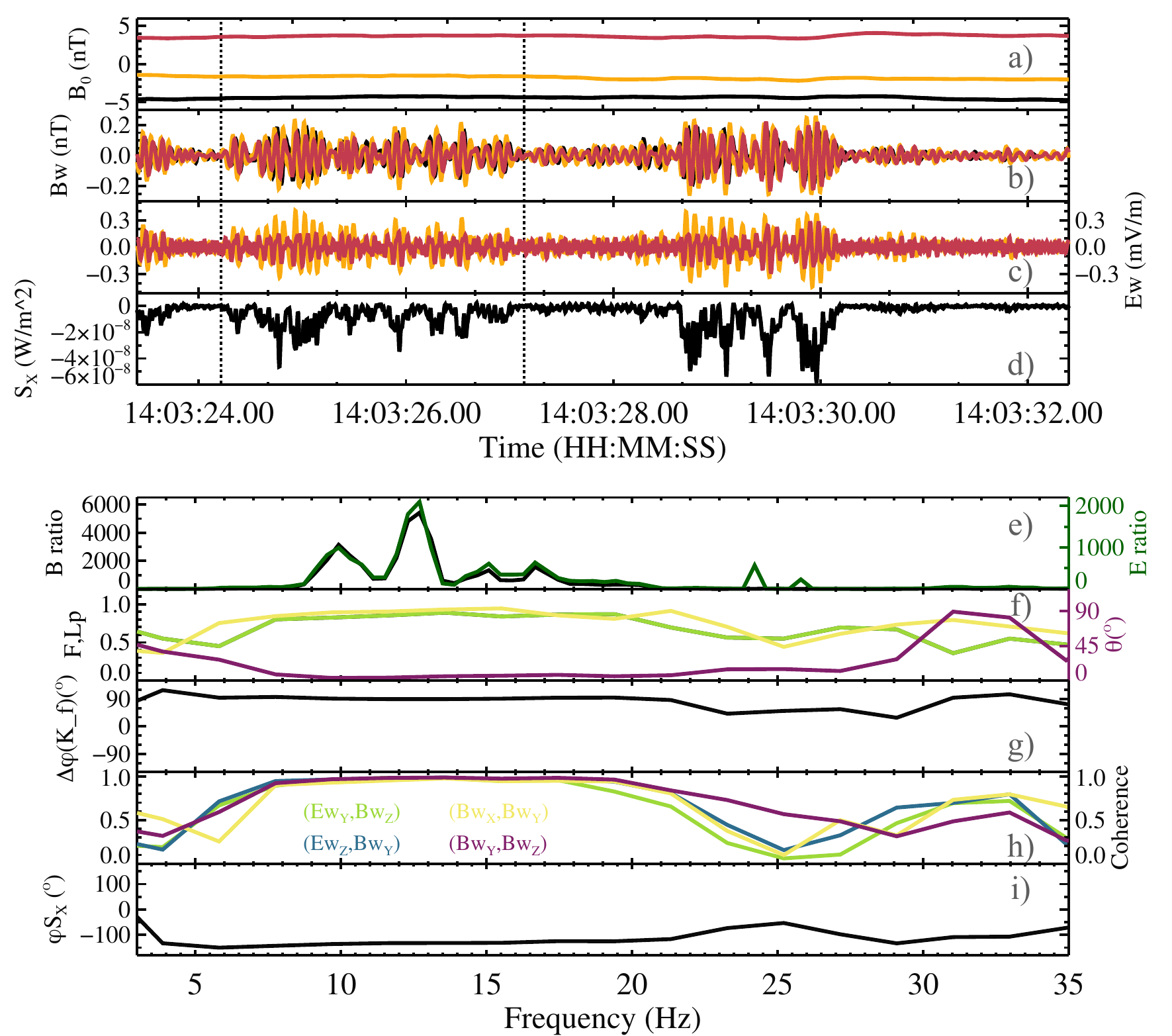}
      \caption{
     Analysis of whistler waves detected by Solar Orbiter RPW on 2020 July, 22nd around 14:03. 
     \textit{Panels a) to d):} The X component is in black, Y is in red, and Z is in yellow (spacecraft reference frame).
     \textit{Panel a):} Background magnetic field $B_{\rm 0}$.
     \textit{Panels b) and c):}
     Waveforms of the magnetic $B{\rm w}$ and electric $E{\rm w}$ fields. 
     \textit{Panel d):}
     X component of the Poynting vector ($S_{\rm X}$). 
     For \textit{Panels c) and d)} we used an effective length of the electric antennas equal to 6 $\si{\meter}$.
     \textit{Panels e) to i):} Analysis of the wave packet delimited by vertical dotted lines.  
     \textit{Panel e):} Ratio of the magnetic power spectrum to the median power spectrum of the day (black) and the same for the electric field (green, with the Y and Z components only). 
      \textit{Panel f):} Polarization (Lp in orange), planarity (F in black), and propagation angle ($\theta$ in purple). 
      \textit{Panel g):} Phase shift between $B{\rm w}_{\rm x}$ and $B{\rm w}_{\rm y}$ (in the wave reference frame with  $\vec{k}=k \vec{z}$, see Section \ref{Data and Analysis}).
      \textit{Panel h):} Spectral coherence between different components. 
      \textit{Panel i):} Phase of the X component of the Poynting vector $\varphi_{S_{\rm X}}$.
      }
         \label{contexte}
   \end{figure}

This statistical analysis includes several improvements over previous works. 
Most other statistical studies used only the power spectrum to detect whistlers without checking the polarization properties \citep{lacombe_whistler_2014,tong_statistical_2019,jagarlamudi_whistler_2020,jagarlamudi_whistler_2021}. 
In addition, the joint analysis of the magnetic and electric fields makes it possible to determine the direction of propagation, which is the most crucial parameter for understanding the wave interactions with electrons. This also allows us to take into account the Doppler effect and to determine the wave frequency in the plasma frame. This joint statistical analysis of magnetic and electric fields was not performed in most of the previous studies of whistler waves in the solar wind \citep{lacombe_whistler_2014,tong_statistical_2019,jagarlamudi_whistler_2020,jagarlamudi_whistler_2021,cattell_narrowband_2020,cattell_parker_2021,cattell_parker_2022,froment_whistler_2023}. Finally, the analysis of the waveforms allows us to determine precisely the parameters of the wave packets (e.g., amplitude, duration). \\
Figure \ref{contexte} shows a typical example of whistler wave activity, detected by Solar Orbiter RPW on 2020 July, 22th. 
 The spectral energy content is present between approx. 8 and 21 $\si{\hertz}$ (Figure \ref{contexte}e). In this frequency range, the polarization (Lp), planarity (F), and coherence between the different components are close to 1 (Figures \ref{contexte}f and \ref{contexte}h).
The phase of the radial component of the Poynting vector, $\varphi_{S_{\rm X}}$, is close to -130 $\si{\degree}$ (Figure \ref{contexte}i). Taking into account the instrumental phase shift correction of -50$^\circ$, this indicates an anti-sunward propagation that is consistent with the negative sign of the X component of the Poynting vector (Figure \ref{contexte}d). Taking into account this anti-sunward propagation, we notice that the propagation angle $\theta$ is small ($\leq 20 ^\circ$, on the frequency range considered), which indicates a quasi-parallel propagation (Figure \ref{contexte}f). Since we have an anti-sunward propagation, in the background magnetic field direction, the phase shift of 90$\si{\degree}$ (Figures \ref{contexte}g) indicates a right-handed circular polarization in both the spacecraft and plasma frame, as expected.
 
\subsection{Results}
\subsubsection{Whistler waves occurence and polarization properties}
\label{Statistical analysis} 

\begin{figure}
   \centering
   \includegraphics[scale=0.48]{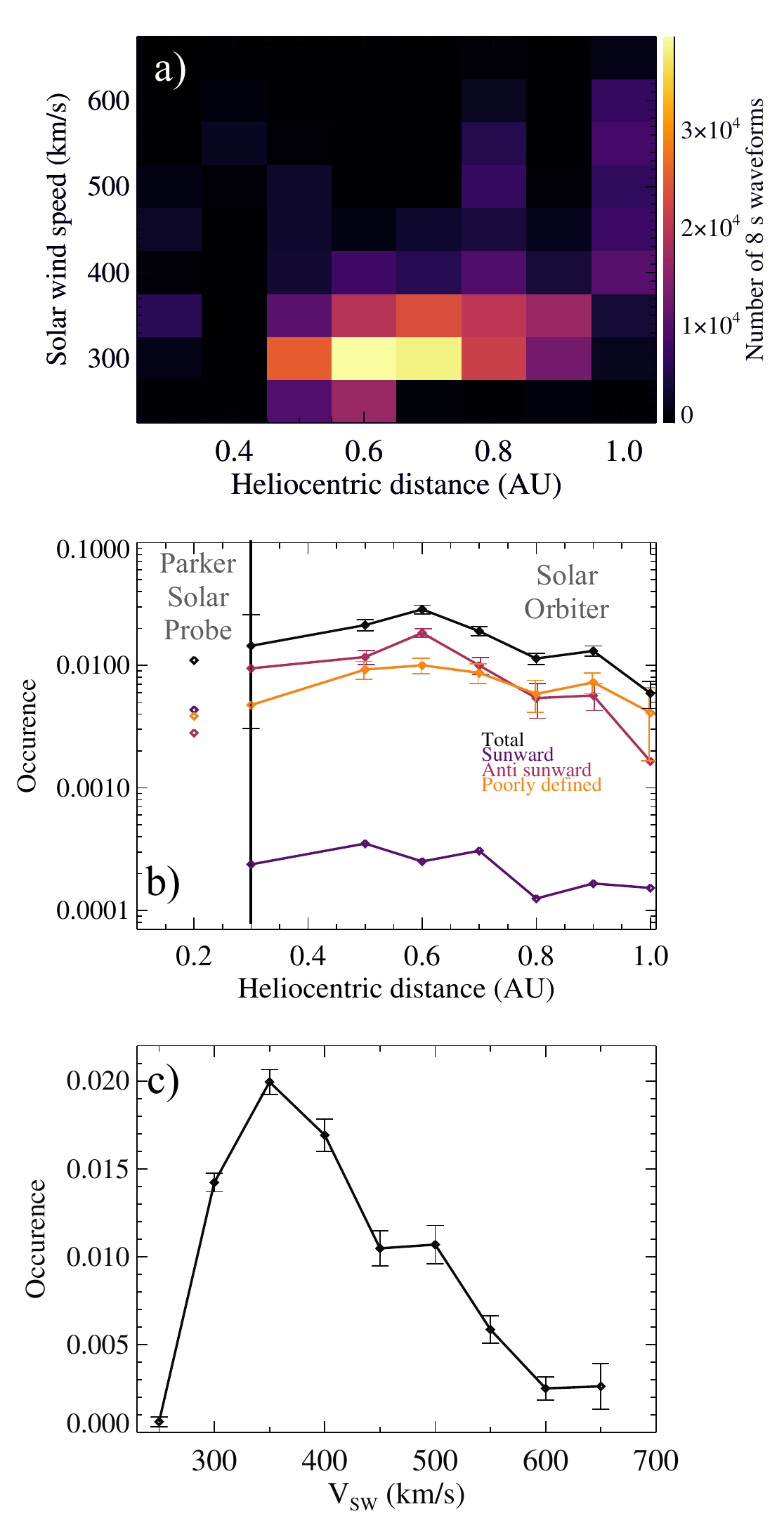}
      \caption{
     \textit{Panel a):} Number of analyzed 8s-waveforms with Solar Orbiter as a function of heliocentric distance and solar wind speed. 
     \textit{Panel b):} Whistler wave occurrence as a function of heliocentric distance for sunward, anti-sunward, and poorly defined propagation direction. 
     When, due to low statistics, the $95 \%$ confidence intervals (calculated assuming a normal distribution) are of the order of magnitude of values, they are not shown. 
     \textit{Panel c):} Occurrence as a function of the solar wind speed. 
     }
         \label{occurence}
   \end{figure}

Figure \ref{occurence} gives an overview of the observations and occurrence of whistler waves. The occurrence is defined as the summed duration of the wave packets divided by the total observation time. Its absolute value cannot, therefore, be compared directly with occurrences computed with data of lower temporal resolution \citep{lacombe_whistler_2014,jagarlamudi_whistler_2020,jagarlamudi_whistler_2021,froment_whistler_2023}. Figure \ref{occurence}a is made using Solar Orbiter data only, between 0.3 and 1 AU and when the solar wind speed data were available. 
Figures \ref{occurence}b and \ref{occurence}c are obtained with both Solar Orbiter and Parker Solar Probe observations. The occurrence for Parker Solar Probe is computed using burst waveforms only as the continuous waveforms have a too low acquisition frequency and miss many whistler waves.
Nevertheless, the occurrence calculated with Parker Solar Probe data is biased because burst waveforms are triggered to detect intermittent waves in the electric field \citep{bale_fields_2016,malaspina_digital_2016}. It is therefore likely that the occurrence of whistler waves is overestimated at 0.2 AU. On the other hand, it allows us to give an estimation of the proportion of sunward and anti-sunward waves.  \\
Figure \ref{occurence}a shows that most Solar Orbiter observations were made in a slow wind ($90\%$ with $V_{\rm SW}$ $\leq$ 450 $\si{\kilo \meter} / \si{\second}$) and between 0.45 and 0.75 AU. We analyze in total 2673 hours of observations with Solar Orbiter and 68 hours of observations with Parker Solar Probe. \\
More than 110,000 whistler wave packets are detected and analyzed, which is the largest statistical study of whistler waves in the solar wind to date. 
Because of the lower statistics with Parker Solar Probe (232 wave packets) and between 0.3 and 0.5 AU with Solar Orbiter (900 wave packets), we concentrate in some figures only on distances between 0.5 and 1 AU. \\
Figure \ref{occurence}b shows that the occurrence rate increases from 1 to 0.6 AU and decreases from 0.6 to 0.2 AU, and that anti-sunward whistler waves are largely dominant above 0.3 AU.  The increase of the occurrence from 1 to 0.6 AU is in agreement with the observations of \cite{kretzschmar_whistler_2021} but in disagreement with the results of \cite{jagarlamudi_whistler_2020} (using magnetic spectra from the HELIOS mission).
There seems to be a slow increase in the number of sunward whistlers between 1 and 0.3 AU but more statistics are needed to verify this trend. At 0.2 AU the occurrence of sunward whistlers is about the same as that of anti-sunward whistlers, but again more statistics are necessary. \cite{kretzschmar_whistler_2021} also found a huge majority of anti-sunward whistler waves between 0.5 and 1 AU. The evolution of the occurrence with distance is discussed in more details in Section \ref{discussion}.  \\ 
Finally, there is an important decrease in the occurrence with increasing solar wind speed above 350 $\si{\kilo \meter} / \si{\second}$
(0.02 à 350 $\si{\kilo \meter} / \si{\second}$ and 0.0025 at 650 $\si{\kilo \meter} / \si{\second}$, Figure \ref{occurence}c). We also note a very fast decrease in the occurrence in very slow wind (0.001 at 250 $\si{\kilo \meter} / \si{\second}$). These results are in good agreement with the study of \cite{tong_statistical_2019} (using Artemis data). A decrease in the occurrence with increasing solar wind speed (above 300 $\si{\kilo \meter} / \si{\second}$) was also found by \cite{jagarlamudi_whistler_2020} and \cite{jagarlamudi_whistler_2021}. Moreover, observations of \cite{cattell_parker_2022} also suggest a decrease in occurrence with the solar wind speed. Note that \cite{kretzschmar_whistler_2021} did not study the occurrence as a function of solar wind speed.
\cite{jagarlamudi_whistler_2020} showed that the lower turbulence level and the lower Doppler effect in the slow solar wind were not sufficient to explain the increase in the occurrence. The temperature anisotropy of the core and of the halo being more important in a slow solar wind, the generation mechanisms for whistler waves (heat flux and temperature anisotropy instabilities) are more favorable. The low occurrence at 250 $\si{\kilo \meter} / \si{\second}$ (representing around 5\% of observations) still needs to be explained. It should be noted that SWA/PAS solar wind speed data are less reliable in very slow solar winds ($< 300 \si{\kilo \meter} / \si{\second}$). However, this has no impact on wave detection with SCM.


    \begin{figure}
   \centering
   \includegraphics[width=\linewidth]{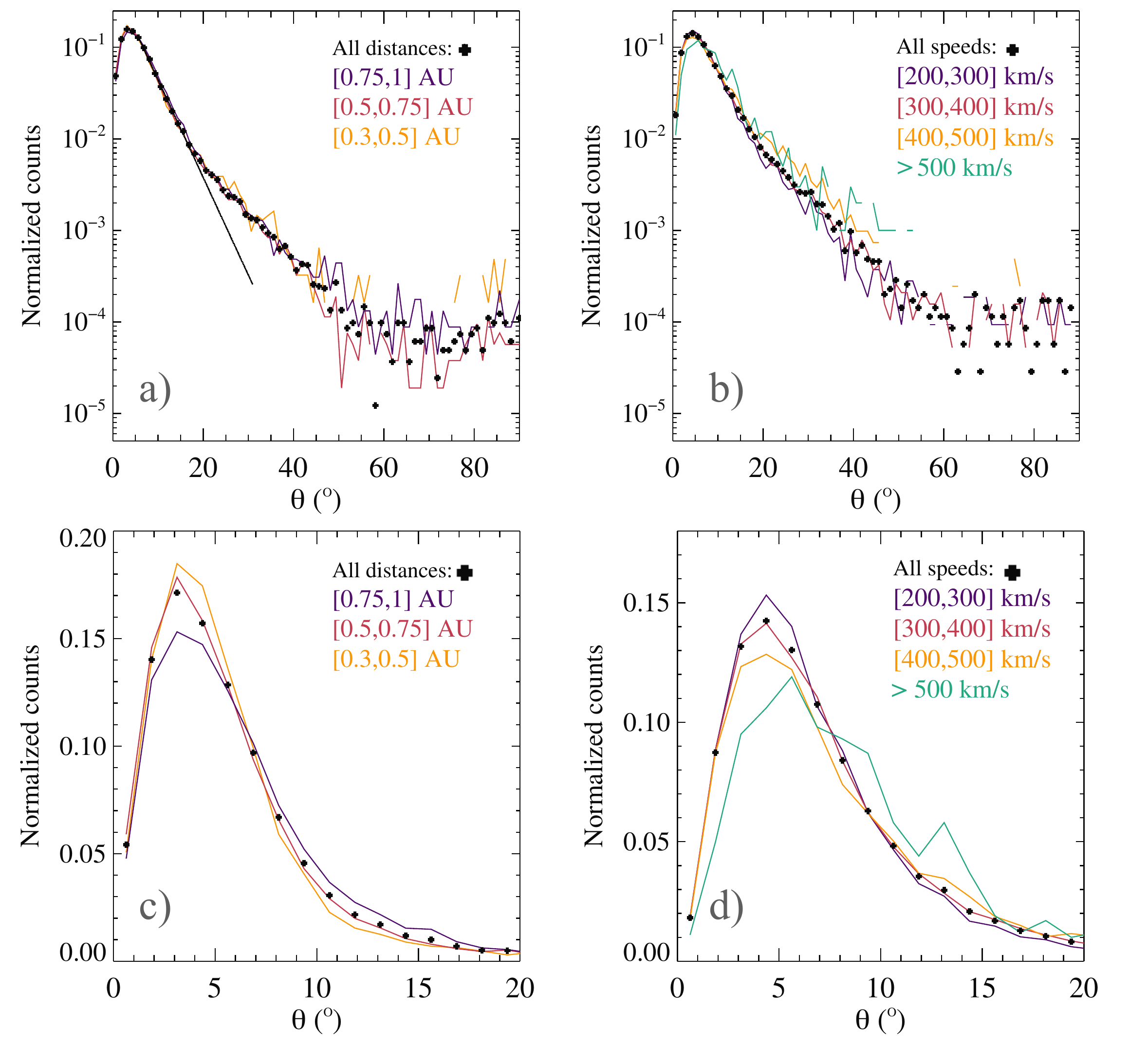}
      \caption{
     \textit{Panel a):} Normalized histogram of the number of whistler waves as a function of the propagation angle for different heliocentric distances.  
     \textit{Panel b):} Normalized histogram of the number of whistler waves as a function of the propagation angle for different solar wind speeds.
     \textit{Panel c):} Same as \textit{panel a)} zoomed between 0 and 20$^\circ$.
      \textit{Panel d):} Same as \textit{panel b)} zoomed between 0 and 20$^\circ$. 
     }
         \label{angle_propagation}
   \end{figure}

\begin{figure*}
   \centering
   \includegraphics[width=0.90\linewidth]{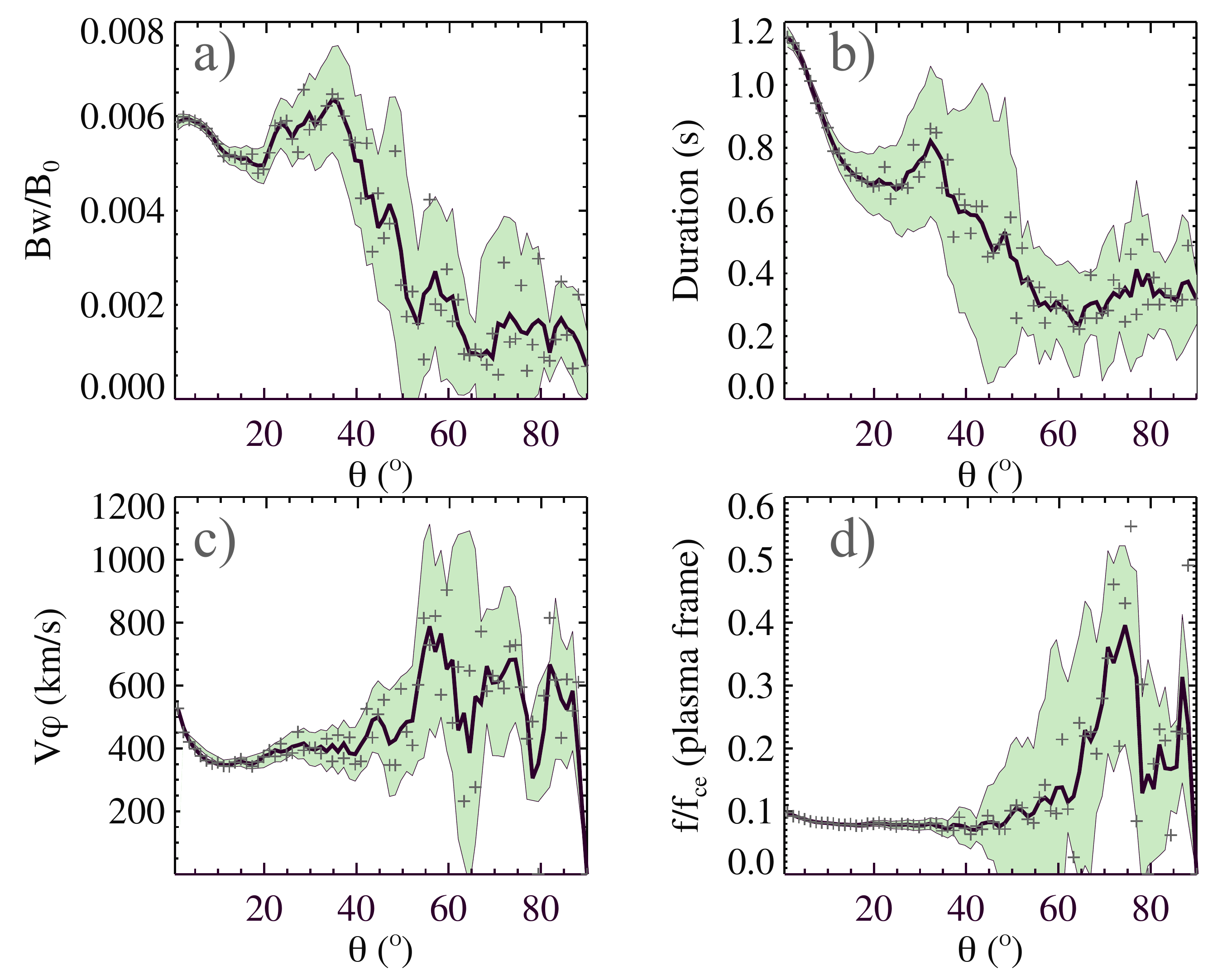}
      \caption{
       Median value (crosses) of several whistler wave parameters as a function of the propagation angle. The black lines show the smoothed values and the smoothed 95\% error bars (assuming a normal distribution) are shown in green. 
       \textit{Panel a):} Wave amplitude normalized to the background magnetic field amplitude. \textit{Panel b):} Duration of the wave packets.
       \textit{Panel c):} Theoretical phase velocity ($v_{\varphi}$) in the plasma frame. 
       \textit{Panel d):} Frequency in the plasma frame normalized by the local electron cyclotron frequency.       
    }
         \label{theta_var}
   \end{figure*}

Figure \ref{angle_propagation} shows the distribution of the wave propagation angle of anti-sunward whistlers measured with Solar Orbiter and its variations with heliocentric distance (Figures \ref{angle_propagation}a and \ref{angle_propagation}c) and solar wind speed (Figures \ref{angle_propagation}b and \ref{angle_propagation}d). Most whistler waves are found to be quasi-aligned with the magnetic field. Indeed, for all distances and speeds, the distribution of the propagation angle peaks between 3 and 7$^{\circ}$. Moreover, only 7.5 $\%$ of the waves have a propagation angle larger than $15^\circ$ and 0.5 $\%$ have an angle of propagation greater than 45$^\circ$. These results are in good agreement with the studies of \cite{lacombe_whistler_2014}, \cite{tong_statistical_2019}, and, \cite{kretzschmar_whistler_2021} that also found a majority of quasi-aligned whistlers between 0.5 and 1 AU. Below $15^\circ$, the distribution can be explained by considering an instability that has a maximum growth rate at 0$^\circ$, which is the case of the WHFI, the TAI, or the sunward suprathermal deficit instability. The departure from 0$^\circ$ can be explained with a geometrical effect due to the curvature of the field lines \citep{agapitov_statistics_2013}. 
This explanation is consistent with the slight increase in propagation angle with the heliocentric distance (already observed by \cite{kretzschmar_whistler_2021} and noticeable in Figure \ref{angle_propagation}c) and with the solar wind speed (Figure \ref{angle_propagation}d). Indeed, the field lines are less curved close to the Sun (considering a simple Parker spiral) and whistlers propagate over greater distances as solar wind speed increases. More measurements within the fast solar wind and closer to the Sun would allow us to confirm this point more rigorously. \\
Between $15^\circ$ and up to $45^\circ$, one can notice a slight change in the slope of the distribution with respect to smaller angles; this is accompanied by other changes that we describe below together with Figure \ref{theta_var}. \\
Above $45^\circ$, very few waves are detected, and a detailed analysis of these cases (not carried out here) would be needed to determine their physical or non-physical origins. \\
At 0.2 AU (distribution is not shown because the number of cases is too small), we also find a majority of quasi-aligned whistler waves with, however, a higher percentage of waves propagating obliquely ($\sim 4 \%$ with $\theta \geq 45^\circ$). This percentage is comparable with the results of \cite{froment_whistler_2023} (3$\%$ with $\theta\geq 45^\circ$) using cross spectra of Parker Solar Probe. This higher percentage can be explained by the fact that, at 0.2 AU, whistlers are often associated with structures such as magnetic dips or switchbacks and therefore propagate in a highly inhomogeneous background magnetic field \citep{agapitov_sunward-propagating_2020,froment_whistler_2023,karbashewski_whistler_2023}. \\


Figure \ref{theta_var} shows the evolution of the median value of several wave parameters with the angle of propagation. To calculate the theoretical phase velocity we use the wave vector from the cold plasma dispersion equation \citep{lyons_pitch_1974}.
Below $15^\circ$, we note a decrease of some parameters with the propagation angle (Figures \ref{theta_var}a, \ref{theta_var}b and \ref{theta_var}c). This may be due to the fact that waves with larger propagation angles are generated at greater distances from the spacecraft. These changes would in this case be due to wave propagation. 
However, we note that these parameters increase slightly (or there is a plateau) from 15$^\circ$ to around 45$^\circ$. This is clearly visible in Figures \ref{theta_var}a, \ref{theta_var}b and \ref{theta_var}c, even though the error bars are large due to the low number of waves verifying $\theta \ge 15^\circ$. On the other hand, the frequency in the plasma frame divided by the electron cyclotron frequency varies only slightly with the propagation angle (Figure \ref{theta_var}d).
Observations of Figures \ref{theta_var}a to \ref{theta_var}c, coupled with the change in slope of the distribution (noted in Figures \ref{angle_propagation}a and \ref{angle_propagation}b), could be the signature of different conditions (e.g., particularly inhomogeneous magnetic field like in dips and switchbacks) and/or mechanisms (instability having a maximum growth rate between $15^\circ$ and $45^\circ$) generating these waves. Further investigations are needed. \\


    \begin{figure*}
   \centering
   \includegraphics[width=0.73\linewidth]{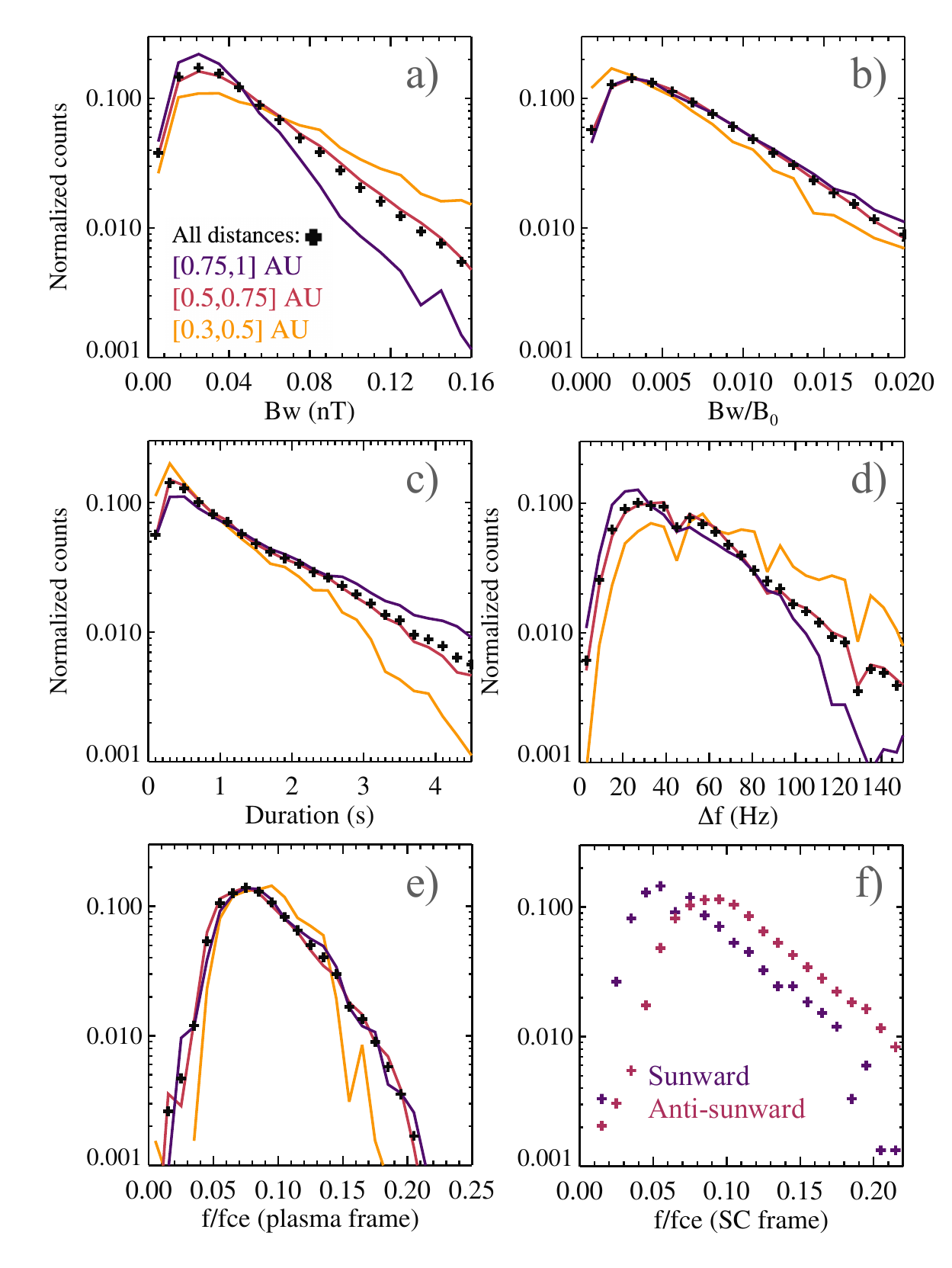}
      \caption{
        \textit{Panels a) to e)}: Normalized histograms for different wave parameters as a function of heliocentric distance. The wave parameters are as follows: 
       \textit{Panel a):} Wave amplitude. \textit{Panel b):} Wave amplitude normalized by the the background magnetic field. \textit{Panel c):} Duration of the wave packets. \textit{Panel d):} Frequency width. \textit{Panel e):} Frequency in the plasma frame normalized by the local electron cyclotron frequency. \textit{Panel f):} Histogram of the frequency in the spacecraft frame normalized by the electron cyclotron frequency for sunward (in purle) and anti-sunward whistler waves (in pink).
    }
         \label{6parameters}
   \end{figure*}

Figure \ref{6parameters} shows the distribution of various wave parameters with heliocentric distance. It can be seen in Figure \ref{6parameters}a that the amplitude of the waves decreases with the heliocentric distance. This is expected since the amplitude of the background magnetic field decreases as $\sim r^{-2}$ with distance (at the distances we consider). Nevertheless, there may be a slight increase in the normalized amplitude of the fluctuations with the heliocentric distance (Figure \ref{6parameters}b).
This increase may be caused by the fact that both the temperature anisotropy \citep{jagarlamudi_whistler_2020,stverak_electron_2015} and
 the plasma $\beta_{\rm e}$ increase with distance. Indeed, \cite{tong_statistical_2019} have shown that these two parameters control the ratio $\frac{B{\rm w}}{B_{\rm 0}}$. Longer wave packet durations are observed far from the Sun (Figure \ref{6parameters}c). This is probably caused by the decreasing phase velocity at larger heliocentric distances and by the fact that the characteristic spatial scales of the plasma are much larger far from the Sun. 
The frequency width decreases with the Heliocentric distance (Figure \ref{6parameters}d). The trends observed in Figures \ref{6parameters}a, \ref{6parameters}b and \ref{6parameters}d confirm what was already observed by \cite{kretzschmar_whistler_2021}. The ratio $f/f_{\rm ce}$ does not clearly depend on the heliocentric distance (Figure \ref{6parameters}e) in the plasma frame, while this ratio seems to increase with distance in the spacecraft frame \citep{kretzschmar_whistler_2021}. This can be explained by a more important role of the Doppler shift far from the Sun because of the smaller phase velocity. 
Figure \ref{6parameters}f shows that $f/f_{\rm ce}$ (in the spacecraft frame) is lower for sunward whistlers than for anti-sunward ones. This is explained by the Doppler effect and gives a good indication that the determination of the propagation direction (sunward or anti-sunward) is correct. 
The observed amplitudes and frequencies are in agreement with previous observations of small amplitude quasi-parallel whistler waves in the solar wind \citep{lacombe_whistler_2014,tong_statistical_2019,chust_observations_2021,kretzschmar_whistler_2021,froment_whistler_2023}, and are consistent with whistler heat flux instability simulations \citep{kuzichev_nonlinear_2019,lopez_particle--cell_2019}.


\subsubsection{Are sunward whistler waves counter-propagating with respect to Strahl electrons?}
\label{Focus on sunward whistler wave}

\begin{figure*}
   \centering
   \includegraphics[width=0.90\linewidth]{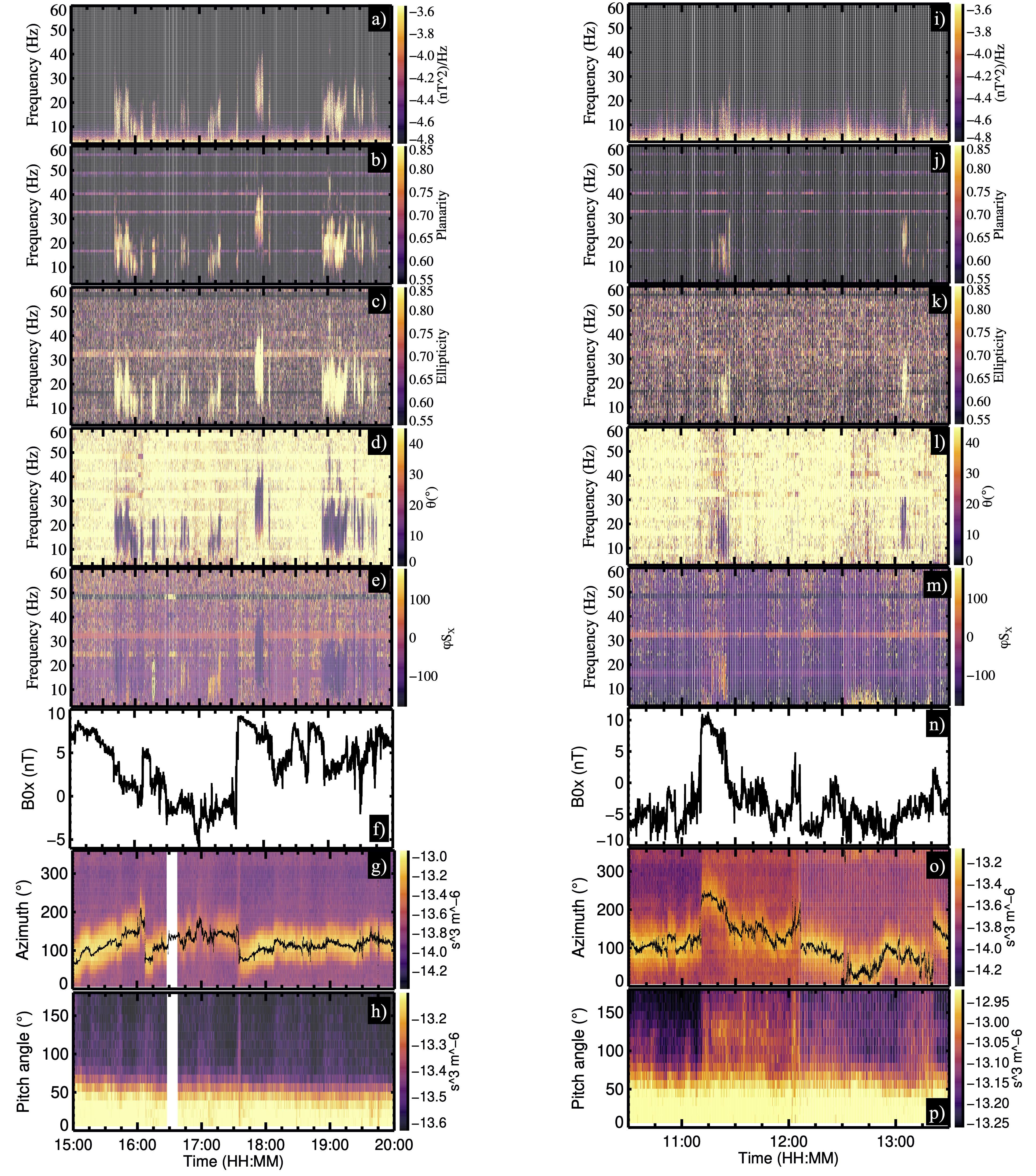}
      \caption{Observations of whistler waves activities with Solar Orbiter on 2021/08/15 on the left and 2021/10/08 on the right. 
\textit{Panels a) to  e) and i) to m):} Spectrograms of several wave parameters calculated from CWF with: 
 \textit{Panels a) and i):} Power spectral density. \textit{Panels b) and j):} Planarity. \textit{Panels c) and k):} Ellipticity. \textit{Panels d) and l):} Propagation angle.  \textit{Panels e) and m):} $\varphi S_{X}$. 
 \textit{Panels f) and n):} X components of the background magnetic field. 
 \textit{Panels g) and o):} Velocity distribution functions (summed over energies between 100 and 1000 eV and over all elevations) as a function of time and azimuth angle of EAS 1 with the azimuth of the background magnetic field in black (in the EAS 1 reference frame).
 \textit{Panels h) and p):} Pitch angle distribution (summed over energies between 100 and 1000 eV) using EAS 1 and EAS 2 data. 
 }
         \label{SB}
   \end{figure*}

We have noted Figure \ref{occurence}b the interesting fact that the proportion of sunward whistler waves seems to increase when getting closer to the Sun. However, since 1) the number of detected sunward whistler waves is always quite low between 0.3 and 1 AU and 2) there is an instrumental phase shift between the magnetic and electric fields on Solar Orbiter, we pay special attention to these sunward waves in this section. Figure \ref{SB} shows two examples of sunward whistler waves activity detected with Solar Orbiter. 
Figures \ref{SB}a to \ref{SB}d and \ref{SB}i to \ref{SB}l are typical of whistler waves observed with Solar Orbiter, with a planarity and an ellipticity close to 1 and a quasi-parallel propagation. 
Figures \ref{SB}e and \ref{SB}m show the phase of the X component of the Poynting vector ($\varphi_{S_{\rm x}}$) that is used to determine the radial direction of propagation (sunward or anti-sunward) of the waves. Taking into account the instrumental phase shift, as explained in Section \ref{Detection and characterisation}, $-180^\circ \leq \varphi_{S_{\rm x}} \leq -80^\circ$  indicates a wave propagating anti-sunward, while  $0^\circ \leq \varphi_{S_{\rm x}} \leq 100^\circ$ indicates a wave propagating sunward. Two sunward propagating cases are clearly visible between 16:30 and 17:30 on the 2021 August, 15th (Figure \ref{SB}e) and between 11:00 and 11:30 on the 2021 October, 8th (Figure \ref{SB}m). The other observed wave packets are propagating anti-sunward or are poorly defined. 
The surprising feature is that these sunward propagating waves occur during switchbacks with a change of sign of the radial component of the background magnetic, such that they are still aligned with the Strahl. This can be seen by observing the reversal of the radial component (X in solar Orbiter reference frame) of the background magnetic field at the time of the observed sunward waves (positive to negative in Figure \ref{SB}f, negative to positive in Figure \ref{SB}n), while the Strahl stays aligned with the magnetic field (Figures \ref{SB}g, \ref{SB}o, \ref{SB}h and \ref{SB}p).  \\



\begin{figure}
   \centering
   \includegraphics[width=0.8\linewidth]{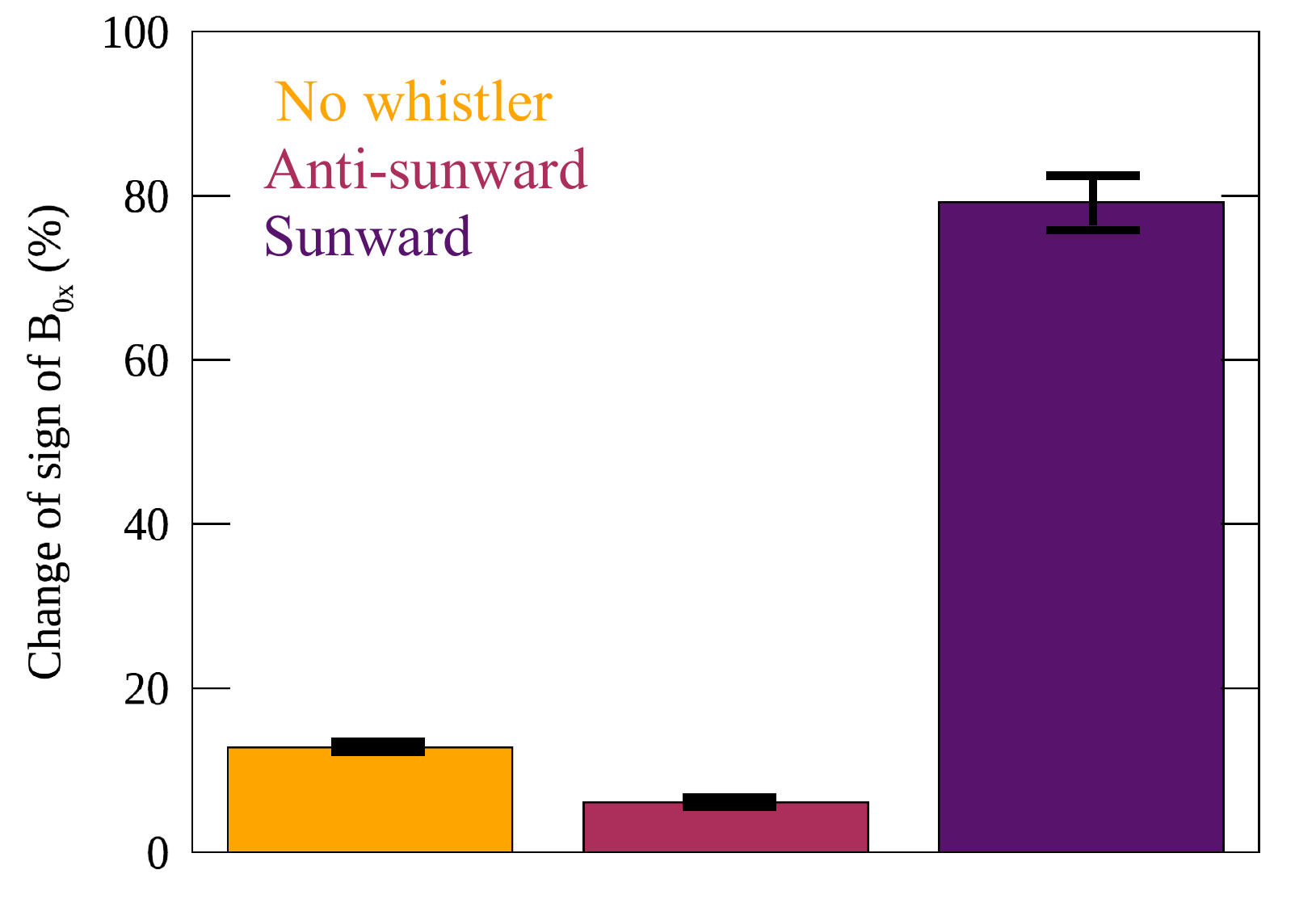}
      \caption{
       Percentage of cases in which the sign of the radial component of the background magnetic field during the measurement is not the median sign of this component over a 24-hour period centered on the measurement. To produce this figure, we use only Solar Orbiter data.
    }
         \label{Sunward_or_not}
   \end{figure}

We now want to estimate if cases similar to those shown in Figure \ref{SB} represent the majority of sunward cases observed with Solar Orbiter.
Figure \ref{Sunward_or_not} presents the percentage of cases in which the sign of the radial component during the measurement is not the median sign of this component over a period of 24 hours centered on the measurement. Such a change in the sign of the radial component would suggest the presence of switchbacks, as in Figure \ref{SB}. Note that this technique is not perfect since we do not verify that the Strahl changes direction as well; heliospheric current sheets crossing could cause a change of sign of the radial component of the magnetic field as well. Note that the error bars in Figure \ref{Sunward_or_not} are estimated using the binomial distribution and do not take into account possible method errors described above.
The percentage of cases with a change of sign of the radial component is around 13 times higher for sunward waves ($\simeq 80\%$) than for anti-sunward waves ($\simeq 6\%$). Without whistler waves, this percentage is $\simeq 12\%$. This suggests that in the vast majority of cases, sunward waves are indeed detected within switchbacks. Therefore, these "sunward" waves actually propagate in the same direction as the Strahl and do not interact with electrons as counter-streaming waves. To keep this distinction in mind when discussing the impact of waves on the suprathermal electrons, we use the distinctive terms of Strahl-aligned and counter-streaming whistler waves. The $6\%$ of anti-sunward waves with a sign change of the radial component and the 20 $\%$ of sunward waves without a sign change are potentially counter-streaming. On the other hand, visual inspection of these cases shows that the vast majority of them are associated with heliospheric current sheets, biasing the detection method. The vast majority of these cases are therefore also Strahl-aligned.  \\

Conversely and interestingly, we did not observe switchbacks with a change of sign of the radial component of the background magnetic field during sunward whistler wave activity closer to the Sun (i.e., with Parker Solar Probe). These sunward waves are therefore really counter-streaming and can strongly contribute to the diffusion of Strahl electrons. These results do not exclude the possibility of observing sunward waves within switchbacks with a complete reversal of the radial component for other Parker Solar Probe encounters. These cases would therefore be Strahl-aligned. Furthermore, as shown by \cite{agapitov_sunward-propagating_2020}, sunward waves can be located at the boundary of a switchback with a complete reversal of the radial component. For these cases, it is necessary to study the electron distribution function in detail to determine the wave propagation direction with respect to the Strahl.
Finally, we found that the observed suward waves are associated with magnetic dips (bigger than $5\%$) in 80 $\%$ of the cases (compared to only $30\%$ for the Strahl-aligned waves).
The magnetic dips are detected similarly to the method described in \cite{froment_whistler_2023}. In the present paper, we compute:
$(|B_{\rm 0}| - |B_{\rm 0}|_{\rm filt})/|B_{\rm 0}|_{\rm filt}$ and identify the magnetic dips when this quantity is lower than -0.03. $|B_{\rm 0}|$ is the background magnetic field smoothed (sliding average) over 0.05 $\si{\second}$ and $|B_{\rm 0}|_{\rm filt}$ is smoothed over 60 $\si{\second}$.

\subsubsection{Overview of whistler waves detected with Parker Solar Probe and Solar Orbiter between 0.2 and 1 AU}
We now present a quick overview of the results obtained in the previous sections and discuss them.

Between 0.3 and 1 AU, we find a majority of quasi-parallel whistler waves. When $\theta \le 15^\circ$ (i.e., $92.5 \%$ of the cases), the following generation mechanisms can produce waves with the observed properties: WHFI, TAI and sunward suprathermal deficit \citep{tong_whistler_2019,jagarlamudi_whistler_2020,bercic_whistler_2021}. The existence of waves with propagation angles up to $15^\circ$ can be explained by a parallel generation associated with the geometric effect of propagation in a curved magnetic field. The $7\%$ of cases with $\theta \in [15,45]^\circ$ may find their origin in another generation mechanism favoring larger propagation angles; we let this for further studies. There are virtually no waves with propagation angles greater than $45^\circ$ (0.5 $\%$ of the case). Moreover, the waves almost all propagate in the direction of the Strahl propagation, including within switchbacks, and the waves are therefore Strahl-aligned. \\
The quasi-absence of counter-streaming whistler waves beyond 0.3 AU is an important finding. Indeed, \cite{vasko_quasi-parallel_2020} predicted the existence of counter-streaming waves with small frequencies and amplitude and emphasized their potential importance in diffusing the Strahl. These waves might be at too small frequencies and amplitudes to be detected by the Search-Coil Magnetometer of Solar Orbiter. We can note, however, that \cite{khotyaintsev_density_2021} analyzed the low-frequency magnetic field measured by MAG as well as the electric potential from RPW and found only proton-band electromagnetic ion cyclotron (PB-EMIC) waves (and no fast-magnetosonic whistler waves). The absence or the very few number of counter-streaming whistler waves far from the Sun, with respect to observations at 0.2 AU, can probably be explained both by the rarity of magnetic dips larger than $5\%$ beyond 0.3 AU, and by the evolution of the distribution function. \\

Similarly, the fact that there are very few oblique whistlers at all the covered distances is an important result for the diffusion of electrons. Oblique waves ($\sim 70^\circ$) were predicted to be produced by fan-like instabilities \citep{vasko_whistler_2019,verscharen_self-induced_2019,micera_particle--cell_2020,micera_role_2021}, that is favored by a high Strahl's density and drift. Our observations close to the Sun and up to solar wind speed of 500 $\si{\kilo \meter}/\si{\second}$ should cover conditions where the Strahl is generally strong \citep{rosenbauer_survey_1977,pilipp_characteristics_1987,stverak_radial_2009,maksimovic_radial_2005,bercic_scattering_2019}. However, we find that whistler waves with $\theta \ge 45^\circ$ constitute only 4$\%$ at 0.2 AU (consistently with \citep{froment_whistler_2023}) and 0.5$\%$ beyond 0.3 AU of observed waves. As mentioned in the introduction, the near-absence of oblique waves had been predicted by \cite{schroeder_stability_2021} and \cite{jeong_stability_2022}.\\
However, we find an important number of counter-streaming waves at 0.2 AU. They are in majority associated with dips in the background magnetic field, which indicates that these structures favor the generation of counter-streaming whistler waves, as proposed by \cite{agapitov_sunward-propagating_2020,karbashewski_whistler_2023,froment_whistler_2023}. Moreover, it was shown by \cite{colomban_reconstruction_2023} using burst data of Parker Solar Probe that these counter-streaming waves are detected at higher frequencies (in the plasma reference frame) than the Strahl-aligned waves. This is in agreement with the mechanism proposed by \cite{karbashewski_whistler_2023}.


Taking into account the results of this section, we expect wave interactions with the Strahl to be rather slow above 0.3 AU (compared to interaction with oblique or counter-streaming waves). At 0.2 AU, on the other hand, counter-streaming waves interact efficiently. In the next section, we compute the diffusion coefficients to quantify the interactions of observed whistler waves with suprathermal electrons.

\section{Diffusion of solar wind electrons by whistler waves}
\label{Diffusion of solar wind electrons by Whistler waves}


We now focus on the quantification of the effect of whistler waves on suprathermal electrons along electron propagation.
The theory and method used to compute the diffusion coefficients are presented in Section \ref{Theory and method}, while Sections \ref{Results for electron diffusion between 0.5 and 1AU} and \ref{Results for electron diffusion at 0.2 AU} present the results. 

\subsection{Theory and method}
\label{Theory and method}
The effect of waves on particles can be calculated in the framework of quasi-linear theory \citep{vedenov_quasi-linear_1963,yakimenko_absorption_1963}. This framework allows us to describe the diffusion of particles in the velocity space due to their interactions with many small amplitude waves with random phases \citep{kennel_velocity_1966}. The amplitude of the self-excited perturbations must be small enough so that the time variation of the distribution function is slow compared to the wave frequencies. Moreover, the growth rate of the waves must be much less than the wave frequencies. This last hypothesis makes it possible to consider only particles that verify the following resonance relation: 
 
 \begin{equation}
  \omega + \frac{n\omega_{ce}}{\gamma_l}=v_{\rm ||}k_{\rm ||}
   \label{resonance}
 \end{equation}
 where $n$ is the order of the resonance, $\gamma_l$ is the Lorentz factor, $v_{||}$ is the parallel (with respect to the background magnetic field) electron velocity and $k$ is the wave vector. \\

One way to estimate the time scale of the electron diffusion in the velocity space is to calculate the diffusion coefficients \citep{lyons_electron_1971,lyons_pitch-angle_1972,lyons_general_1974}. 
These coefficients are computed from the characteristics of the waves and take into account their polarization properties, their amplitudes, and the resonant conditions of wave-particle interactions. \\
 Three diffusion coefficients, $D_{\alpha \alpha}$, $D_{\alpha,p}$, and $D_{pp}$ can be calculated ($\alpha$ is the pitch angle and p is the electron momentum).
Here, we are interested in the interaction of electrons with whistler waves, which are mainly electromagnetic in their reference frame (moving with the wave). Therefore, the perturbations produced by the waves are primarily magnetic. As it is well known, the magnetic field cannot transfer energy to an isolated charge, and therefore the main effect of the waves is angular diffusion. Although energy diffusion is important for the damping/instability of the waves, its effect on the shape of the overall electron velocity distribution function is negligible. This angular diffusion is, therefore, the only one we will consider in the following, it is described by the diffusion coefficient $D_{\alpha \alpha}$. \\
The detailed method to compute the diffusion coefficients is presented in \cite{lyons_pitch_1974}, \cite{horne_resonant_2003}, \cite{glauert_calculation_2005}, and \cite{artemyev_statistical_2012}. In this study, we use the cold plasma whistler waves dispersion relation \citep{lyons_pitch_1974} and the high density approximation. \cite{glauert_calculation_2005} have shown that if $\omega_{pe}/\omega_{ce} \geq 10$ (which is the case in the solar wind), the high-density approximation is well valid for all energies. \\

We first use the diffusion coefficients
to evaluate the angular diffusion caused by an isolated wave packet. We use $D_{\alpha \alpha}$ to calculate the angular diffusion/widening in angular space of the distribution function ($\Delta \alpha$) for a group of resonant particles, which results from the interaction with a wave packet, using: 

\begin{equation}
    \Delta \alpha ^{2} = 2 D_{\alpha \alpha }\tau / p^{2}
    \label{local_diff}
\end{equation}

where $\tau$ is the wave packet duration. In the following $\Delta \alpha / \tau$ is called the local diffusion coefficient and describes the typical angular diffusion that an electron with momentum $p$ undergoes in one second. We calculate the local diffusion coefficients using the mean values of the wave parameters derived in the previous sections (see details in appendix \ref{Annexe B}) for different bins of heliocentric distances, propagation angles, and wave amplitudes. This allows us to study which wave parameters are important for the diffusion. 
 We focus on the diffusion of halo and Strahl electrons by whistler waves propagating either mainly along the Strahl (Strahl-aligned) or against it (counter-streaming). For simplicity, we only consider two energies, 300 eV and 700 eV, that are representative of these suprathermal electron populations, and at which the Strahl diffusion cannot be explained by collisions \citep{boldyrev_kinetic_2019,bercic_interplay_2021}. In the following, we use an anti-sunward magnetic field convention, so that the electrons with a pitch angle between 0 and 90 $^\circ$ (respectively, 90-180$^\circ$) propagate anti-sunward (respectively, sunward) in the solar wind reference frame.

Then, in order to estimate the global impact of whistler waves on the distribution function along
the propagation, we calculate the integrated diffusion coefficients. For each bin presented above, we compute the number of wave packets that an electron encounters on average during its propagation. We assume that the occurrence and characteristics of whistlers encountered by electrons on their journey are identical to those encountered by the satellites.
We describe each wave packet encounter by a "kick widening": $(\Delta
\alpha ^{2})_{i}$, where the index $i$ numerates wave packet. 
Therefore, the total angular deviation, after multiple encounters with different wave packets for the same
group of resonant particles should be evaluated as follows:

\begin{equation}
(\Delta \alpha ^{2})_{total}=\sum\limits_{i}(\Delta \alpha ^{2})_{i}=2\frac{1%
}{p^{2}}\sum\limits_{i}(D_{\alpha \alpha })_{i}\tau _{i}
\label{integrated_diff}
\end{equation}
The total diffusion is achieved when $\sqrt{(\Delta \alpha
^{2})_{total}}=180^\circ $. In the following $(\Delta \alpha)_{total}$ is called the integrated diffusion coefficient. Since this approach is based on diffusion the kick caused by each encounter with a wave packet is assumed to be small. 

 The travel time of the electrons in each bin of heliocentric distance is calculated by dividing the total length of the interplanetary magnetic field (considered to be equal to 1.6 times the radial distance \citep{graham_investigating_2018}) by electron velocity. Electrons velocity is calculated using their energies (the solar wind speed is neglected). Electrons follow the field lines and their rotation around it is considered negligible compared to the total length.
We take the example of an electron with an energy of 300 eV that corresponds to a speed of 15675 $\si{\kilo \meter}/\si{\second}$. Between 0.5 and 0.75, it travels a distance of about $0.25*1.6*1.496*10^{8} = 59840004$ $\si{\kilo \meter}$ and its travel time is therefore of the order of 1h04.
 
 With the chosen anti-sunward magnetic field convention and at the considered energies, small pitch angles ($\lesssim 55$°, depending on the heliocentric distance) indicate Strahl electrons while larger pitch angles indicate halo electrons. 

Below, we first present the local and integrated diffusion coefficients between 0.5 and 1 AU (using Solar Orbiter, Section \ref{Results for electron diffusion between 0.5 and 1AU}), and then near 0.2 AU (using Parker Solar Probe, Section \ref{Results for electron diffusion at 0.2 AU}). 

\subsection{Electron diffusion between 0.5 and 1 AU}
\label{Results for electron diffusion between 0.5 and 1AU}
\begin{figure*}
   \includegraphics[width=0.99\linewidth]{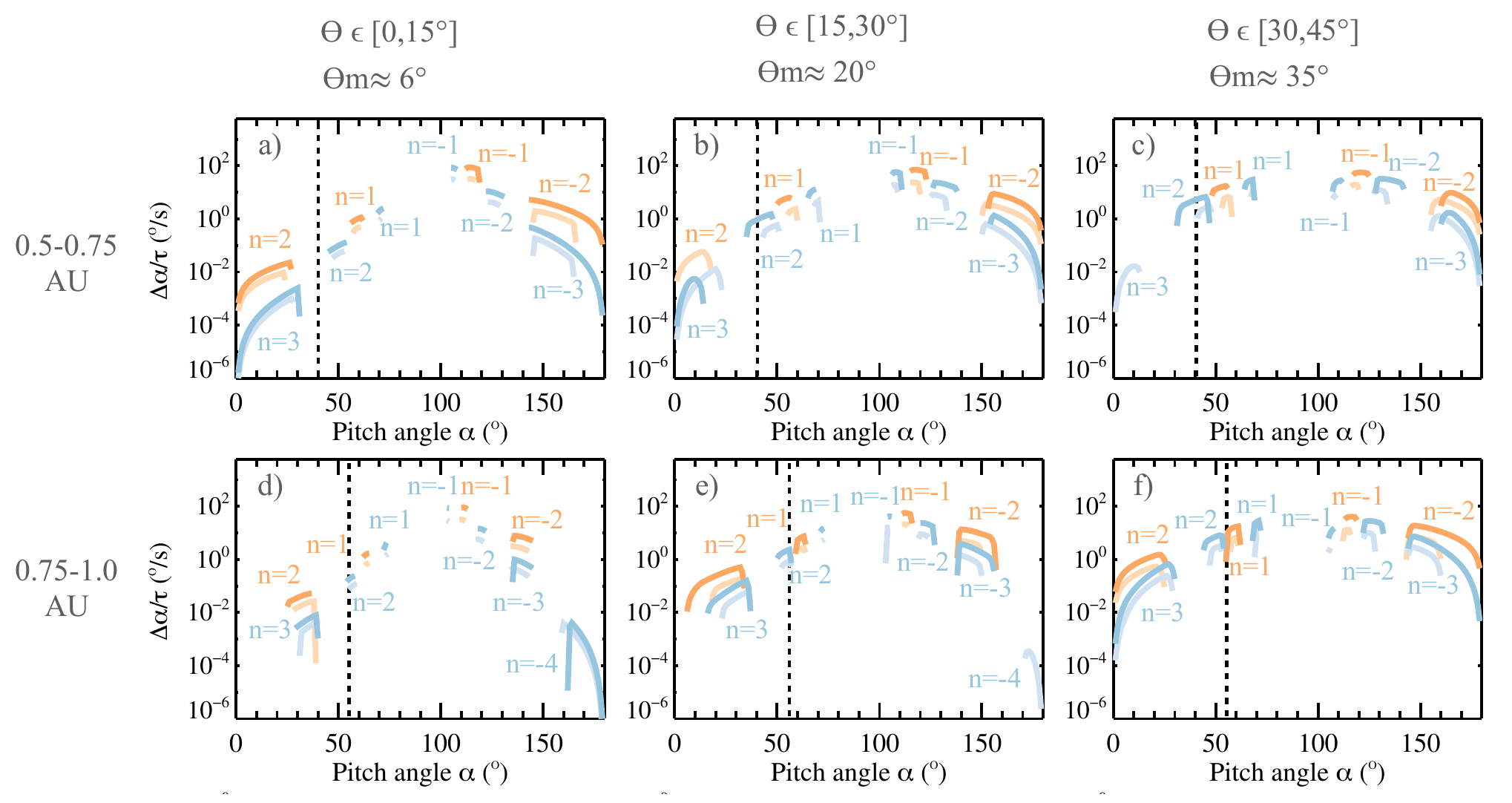}
      \caption{Local diffusion coefficients at two different heliocentric distance (rows) and three wave propagation angle $\theta$ (columns) bins. $\theta_m$ is the approximate mean angle used for the calculation in each propagation angle bin. The orange color (resp., blue) corresponds to an electron energy of 300 eV (resp., 700 eV) in the proton reference frame. The lighter lines show the effect of the 80\% least intense waves while the darker ones are for the 20\% most intense waves. The observed Strahl PAW \citep{bercic_scattering_2019} for each distance bin is indicated by a black vertical line. The resonance orders are indicated. 
        }
         \label{diffusion_local}
   \end{figure*}

\begin{figure*}
   \includegraphics[width=0.99\linewidth]{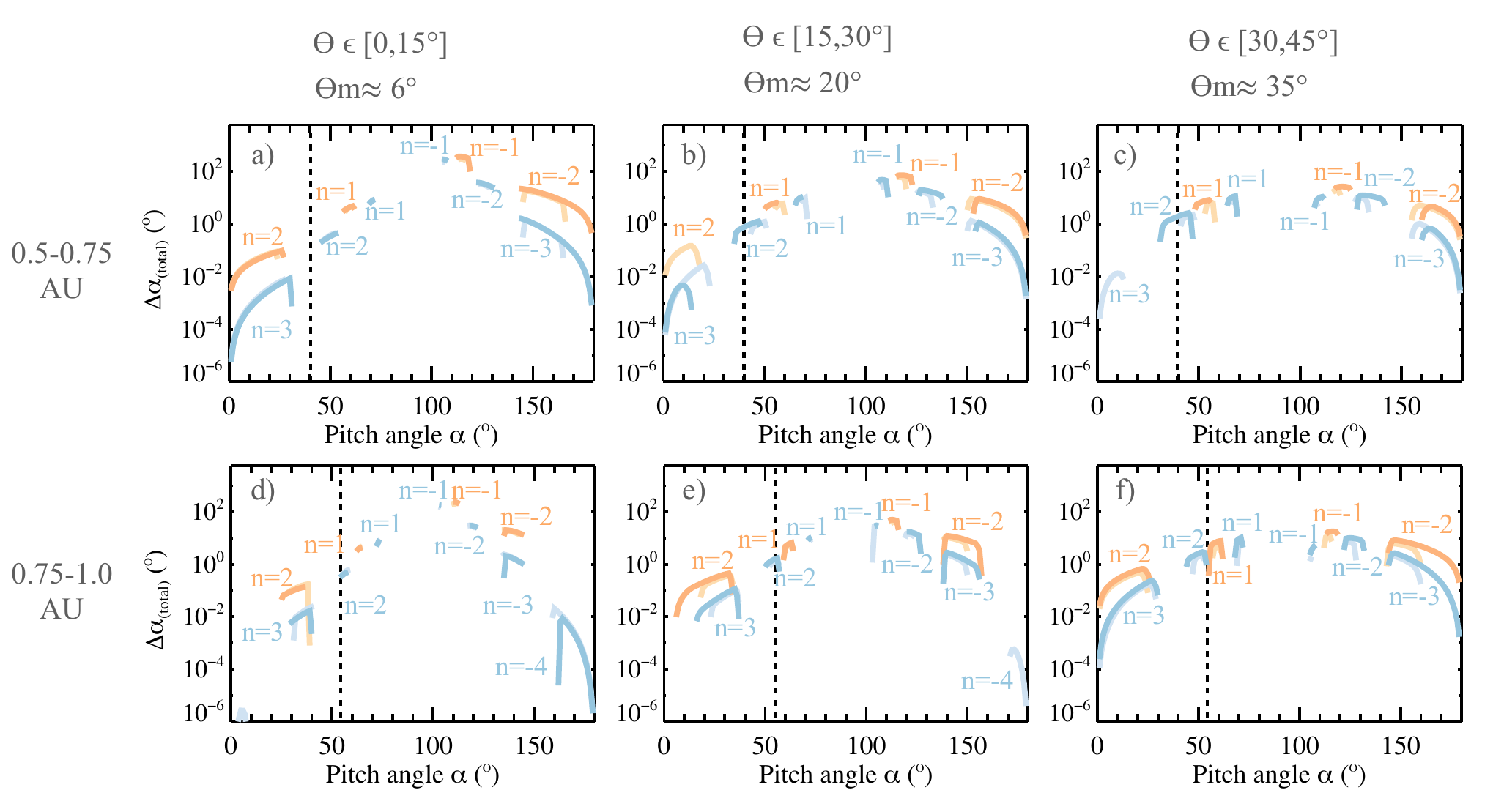}
      \caption{Integrated diffusion coefficients at two heliocentric distance and three propagation angle $\theta$ bins. Legends and notations are similar to Figure \ref{diffusion_local}.
        }
         \label{diffusion_int}
   \end{figure*}

As explained in Section \ref{Focus on sunward whistler wave}, between 0.5 and 1 AU, we detect almost exclusively Strahl-aligned whistler waves.
Figure \ref{diffusion_local} shows the local diffusion coefficients and Figure \ref{diffusion_int} the integrated ones, for this direction of propagation. Three different propagation angle ($[0,15],[15,30],[30,45]^\circ$) and two heliocentric distance ($[0.5,0.75],[0.75,1]$ AU) bins are considered (we find that modifying these bins do not modify the conclusions). Waves with an angle of propagation greater than 45$^\circ$ occur so infrequently that electrons at 700 eV will encounter only about 0.08 waves of this type as they travel $0.25$ AU.  We therefore neglect these waves as their overall effect on the Strahl is negligible, but do not exclude that they can sporadically have an effect on Strahl diffusion.  \\

 Let us first explain some general features of Figures \ref{diffusion_local} and \ref{diffusion_int}. As we work with averaged parameter values for each bin, there is only a specific range of pitch angles that matches the resonance conditions; in reality, the diffusion is caused by the encounter with a multitude of wave packets (with a multitude of parameters) so that electrons at other pitch angles can also be in resonance. To quantify the effect of waves on Strahl electrons we interpolate the diffusion coefficients at the pitch angle corresponding to the Strahl PAW. Strahl PAW values are those obtained by \cite{bercic_scattering_2019} using Helios data (using values of the core electron plasma $\beta_{\rm ec}$ greater than 0.4, coherent with a slow solar wind). Finally, we can note that since whistler activity is intermittent, total diffusion will vary according to the periods of activity encountered. 
 

 
We must also note that in some cases, described below, the wave-particle interaction is found to be very effective such that we are at the limit or outside the applicability of the quasi-linear theory. For such waves, it would be necessary to apply another approach taking into account the effect of nonlinear interactions with a monochromatic solitary wave \cite{karpman_effects_1975}. Nevertheless, as we will see, these cases that concern the diffusion of the halo do not impact the main conclusions of this work. \\ 
Finally, let us remind that Strahl-aligned ($k_{||}>0$) whistler waves can resonate ($n<0$) with electrons verifying $\alpha \le 90^\circ$ with the right polarized part of the wave (for not too oblique waves). They can also resonate ($n>0$) with electrons verifying $\alpha \ge 90^\circ$, mainly with the left polarized part of the wave. It is therefore expected that resonances with $n<0$ and low orders (specifically $n=-1$) are the most efficient. Counter-streaming waves have resonances of the opposite sign. This explains why counter-propagating whistlers, even with a small propagation angle, are very efficient in diffusing the Strahl electrons.

\subsubsection{Local diffusion between 0.5 and 1 AU}
\label{Estimation of diffusion between 0.5 and 1 AU}

We first discuss the local diffusion coefficients and we begin with the role of whistler waves on 300 eV electrons (in orange) between 0.5 and 0.75 AU (first row of Figure \ref{diffusion_local}).
It can be seen in Figure \ref{diffusion_local}a that the diffusion of Strahl electrons by quasi-aligned whistler waves ($\theta \in [0,15]^\circ$), due to the $n=1$ and $n=2$ resonances, is slow. Indeed, even the 20 $\%$ most intense whistler waves (darker curve) can diffuse electrons in pitch angle by about only 0.1 $^\circ/\si{\second}$ at the PAW, which is small considering the typical duration of 1 $\si{\second}$ of a wave packet. 
On the other hand, for pitch angles greater than $90^\circ$, the diffusion is very efficient: up to 100 $^\circ/\si{\second}$ for the $n=-1$ resonance and for the most intense waves. This means that in that last case, the timescale of the variation of the distribution function is not much lower than the wave period: $\sim 5^\circ$/wave period (wave period of the order of 0.05 $\si{\second}$ at these distances). We are, therefore, at the limit of applicability of the quasi-linear theory. Moreover, the approach of diffusion by "kick widening" is also at the limit of applicability since a kick corresponds to a diffusion of the order of $100^\circ$.
We can note in general that the 20\% most intense waves are approximately four times more efficient than the remaining 80\% of the waves.  \\
For $\theta \in [15,30]^\circ$ (Figure \ref{diffusion_local}b), the diffusion of the Strahl is again due to the $n=1$ and $n=2$ resonances and is between 3 and 10 times (depending on the resonance) more effective than for waves with $\theta \in [0,15]^\circ$. The $n=-1$ resonance is again at the limit of applicability of our method. \\
Finally, for $\theta \in [30,45]°$ (Figure \ref{diffusion_local}c), the diffusion of the Strahl becomes much more efficient (about a few degrees per second). However, we remain within the range of applicability of quasi-linear theory: diffusion of $\sim 0.25^\circ/$wave period for n=2.
The diffusion of electrons at $n=-1$ is slightly less efficient than for more aligned waves. Let us recall that the whistler waves with $\theta \in [30,45]$ represent only 1 $\%$ of the total number of waves, so the statistics for the right column are smaller than for the others. \\

Similar conclusions can be drawn for heliocentric distances between 0.75 and 1 AU (2nd row). We note, however, that for quasi-aligned waves (Figures \ref{diffusion_local}d and \ref{diffusion_local}e), the resonances are slightly closer to 90$^\circ$ with respect to Figures \ref{diffusion_local}a and \ref{diffusion_local}b. This is due to the fact that the resonance velocity decreases with distance (not shown), which implies that for the same particle energy the absolute value of $\cos(\alpha)$ must be smaller (Equation \ref{resonance}).  \\
When we consider an electron energy of 700 eV (blue lines), we also notice that the pitch angle of resonance approaches 90$^\circ$ (which is again expected from Equation \ref{resonance}) and that there are resonances at higher harmonics ($n=-4$ for the second row). Finally, using Equation \ref{resonance} it is possible to understand that the range of resonance pitch angles and the difference in resonance pitch angles between the 2 energies increase with $|n|$.

\subsubsection{Integrated diffusion between 0.5 and 1 AU}
We now focus on the integrated diffusion coefficients and start by discussing the total impact of whistler waves on 300 eV electrons between 0.5 and 0.75 AU (first row of Figure \ref{diffusion_int}). At this energy, electrons take about 1h40 to travel a radial distance of 0.25 AU. As the occurrence of the $80\%$ less intense whistler waves with $\theta \in [0,15]^\circ$ is about $1.25\%$ at these distances, the electrons encounter about 75 wave packets. Using Equation \ref{integrated_diff}, we deduce that the total diffusion of Strahl electrons due to these waves is of the order of $0.5^\circ$. The role of the $20\%$ most intense waves is similar. 
Therefore, in spite of the fact that they represent the majority of the observed waves ($92.5 \%$ of the cases), the most aligned whistlers ($\theta \in [0,15]^\circ$) cannot explain the Strahl diffusion (at 300 eV) observed between 0.5 and 0.75 AU (of the order of 10$^\circ$ \citep{bercic_scattering_2019}). 
However, for halo electrons, the $n=-1$ and $n=-2$ resonances provoke highly efficient interactions and contribute effectively to the isotropization of this population.  \\
The total diffusion of 300 eV Strahl electrons by waves with a propagation angle $\theta \in [15,30] ^\circ$ (Figure \ref{diffusion_int}b) is more important because, as discussed in Section \ref{Estimation of diffusion between 0.5 and 1 AU}, there are more effective (up to one order of magnitude). The total diffusion of 300 eV Strahl electrons due to these waves is about $4^\circ$. This diffusion is due to the encounter with an average of 8 wave packets verifying $\theta \in [15,30]^\circ$.
These waves also participate in the isotropization of the halo by the $n<0$ resonances. \\
Finally, whistler waves with propagation angle $\theta \in [30,45] ^\circ$ (Figure \ref{diffusion_int}c) are also important despite their very low occurrence. Indeed, the total diffusion of 300 eV Strahl electrons is of the order of $8^\circ$ and is due to the encounter with an average of 1.5 wave packets.
Their role in the isotropization of the halo with respect to the more aligned waves is negligible. \\

We therefore expect a diffusion of the order of 10$^\circ$ of the 300 eV Strahl electrons between 0.5 and 0.75 AU (by integrating the effect of the waves from all the propagation angle bins) which is consistent with what was observed by \citet{bercic_scattering_2019}. Our method indicates that whistlers with $\theta \in [15,45]^\circ$ seem to be responsible for most of the diffusion of the Strahl electrons (at 300 eV), although they represent only $7\%$ of the waves. In the majority of cases the most important diffusion is at n=2 and not n=1 (resonance pitch angle too large in relation to the Strahl PAW). We are at the limit of applicability of the quasi-linear theory for n=-1 resonance, which leads to an overestimation of the diffusion of the halo. However, our results still show that the waves participate also effectively in the isotropization of the halo electrons, most of which is due to waves having $\theta \le 15^\circ$.  \\
For 700 eV electrons, the conclusions are similar to the case of 300 eV electrons. 
The results between 0.75 and 1 AU (fourth row) are close to those obtained between 0.5 and 0.75 AU, with whistler waves with $\theta \in [15,45]^\circ$ providing a diffusion of the Strahl electrons of the order of 10$^\circ$ at both energies. This is notably facilitated since the PAW is larger at these distances. As we have fewer statistics between 0.3 and 0.5 AU, and in order not to overload the figures, we do not show this distance range in Figures \ref{diffusion_local} and \ref{diffusion_int}. However, as discussed in Section \ref{Statistical analysis}, the occurrence and properties of these waves are similar to those between 0.5 and 1 AU. We can therefore assume that whistler waves are also involved in Strahl scattering between 0.3 and 0.5 AU.

 \subsection{Electron diffusion around 0.2 AU}
 \label{Results for electron diffusion at 0.2 AU}
\begin{figure*}
   \centering
   \includegraphics[width=0.78\linewidth]{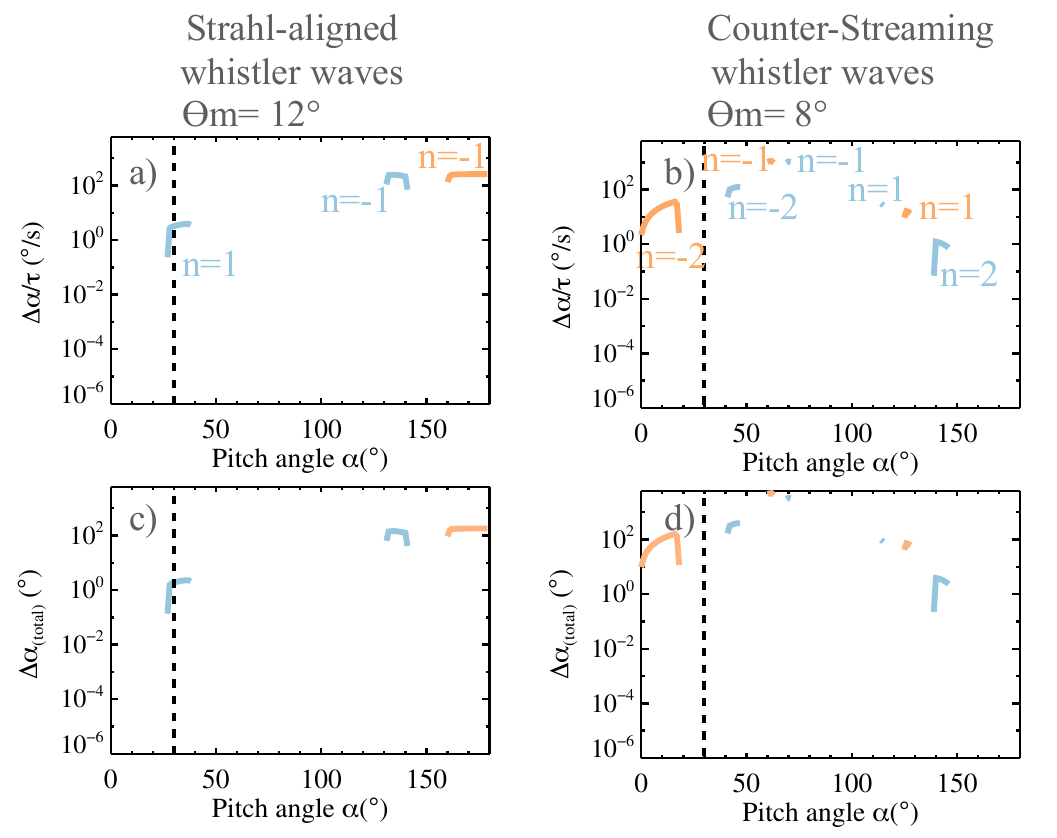}
      \caption{Local (first row) and integrated (second row) diffusion coefficients for Strahl-aligned (left column) and counter-streaming whistler waves (right column), at 0.2 AU. The diffusion coefficients are integrated over 0.05 AU only. Legends and notations are similar to Figure \ref{diffusion_local}.
        }
         \label{diffusion_PSP}
   \end{figure*}

Around 0.2 AU, with Parker Solar Probe observations, we have identified both Strahl-aligned and counter-streaming whistler waves (Section \ref{Focus on sunward whistler wave}). We consider these two cases separately in the following. As there are fewer statistics than with Solar Orbiter, we average over all propagation angles and amplitudes to obtain Figure \ref{diffusion_PSP}. This figure shows the local (first row) and integrated (second row) diffusion coefficients for Strahl-aligned (first column) and counter-streaming (second column) whistler waves. \\

We start by describing the local diffusion of electrons by Strahl-aligned whistler waves (Figure \ref{diffusion_PSP}a). We first note the absence of resonance with $n>0$ for 300 eV electrons (orange color), which prevents the Strahl from being diffused by these waves.
  This absence of resonance seems to be due to a lower $\omega_{\rm pe}/\omega_{\rm ce}$ ratio (and therefore a higher phase speed) for the Strahl-aligned whistlers at these heliocentric distances: $\omega_{\rm pe}/\omega_{\rm ce} \simeq 87$ at 0.2 AU while $\omega_{\rm pe}/\omega_{\rm ce} > 220$ between 0.5 and 1 AU (average values). Nevertheless, since only 30 of these waves were detected, more statistics are necessary to confirm if this is a true effect or if this is caused by low statistics. The resonance at $n=-1$ is close to a pitch angle of $180^\circ$ but gives a diffusion coefficient of the order of 200$^\circ / \si{\second}$. Since the frequency of the waves is more important at this distance (f $\simeq 180 \si{\hertz}$), the quasi-linear theory is still valid for this resonance diffusion of $\sim 1.1^\circ$/wave period). 
 For 700 eV Strahl electrons, only the resonance n=1 (for $\alpha \leq 90^\circ$) occurs, again due to the low $\omega_{\rm pe}/\omega_{\rm ce}$. The average amplitude of these waves being important (0.33 nT, against 0.05 nT between 0.5 and 1 AU) their efficiency is similar to waves with $\theta > 30^\circ$ between 0.5 and 1 AU in spite of the fact that the average angle of propagation is $12^\circ$. 
 
We now focus on the counter-streaming whistler waves (right column). We notice that they are very efficient in diffusing the Strahl (interpolation of local diffusion coefficients gives about a hundred degrees per second at the PAW, Figure \ref{diffusion_PSP}b). The Strahl diffusion is mainly due to the $n=-2$ resonance since the PAW is small at 0.2 AU. The $n=-1$ resonance is out of the range of applicability of quasi-linear theory. 
 
 As explained in Section \ref{Statistical analysis}, the PSP burst mode being triggered, the occurrence of whistler waves is probably overestimated. This therefore induces an overestimation of their global effect (Figures \ref{diffusion_PSP}c and \ref{diffusion_PSP}d). 
 On the other hand, counter-streaming waves are very efficient in diffusing Strahl electrons and account for at least half of all waves. This indicates that they probably play an important role in the Strahl diffusion around 0.2 AU. These waves are probably more important than the oblique waves ($\theta \ge 45^\circ$) in the Strahl scattering since oblique waves account for only 3$\%$ of the whistlers measured \citep{froment_whistler_2023}. 
 
\section{Discussion and conclusions} 
\label{discussion}
\label{On the electron diffusion by Whistler waves}


The main findings of the previous sections can be summarised as follows.

First, around 0.2 AU, both Strahl-aligned and counter-streaming whistler waves are present and can play a role in the Strahl diffusion. However, counter-streaming waves are up to two orders of magnitude more efficient than Strahl-aligned ones. The interaction of these waves with Strahl electrons is mainly due to the $n=-2$ resonance. A highly efficient Strahl electron diffusion process is expected around 0.2 AU to explain the observed increase in PAW in spite of the significant focusing at these distances. 
The integrated effect of the observed waves is difficult to estimate because of the bias in the occurrence rate determined for PSP, due to burst modes being triggered, and because of the low statistics. However, given their very high efficiency, we can assume that counter-streaming waves play an important role in Strahl diffusion around 0.2 AU. \\


Second, between 0.3 and 1 AU, our method suggests that Strahl-aligned whistler waves can explain the observed diffusion of the Strahl electrons. Indeed, our estimate for the total diffusion of the Strahl is of the order of 10$^\circ$ every 0.25 AU, which is consistent with the observations of \cite{bercic_scattering_2019}. 
We found that although whistler waves with $\theta \in [15,45]^\circ$ represent only $7 \%$ of the waves, their efficiency ($\sim 1^\circ / \si{\second}$) is large enough to be responsible for the majority of the Strahl diffusion. Specifically, waves with $\theta \ge 30^\circ$ account for the majority of the diffusion. 
The origin of waves verifying $\theta \in [15,45]^\circ$ is not clearly established and deserves further study. These waves may be due to another instability generating slightly oblique waves or are associated with particular magnetic configurations (e.g. dips, switchbacks). Waves verifying $\theta \ge 45^\circ$, are too few in number and can only have a sporadic impact on the Strahl electrons.
It is therefore important to note that our results suggest that Strahl diffusion is not due to an instability generating oblique whistlers, like the fan instability ($\theta \sim 70^\circ$), as it has been often suggested. It is also interesting to note that, contrary to the results obtained in the radiation belts \citep{artemyev_statistical_2012}, it is not only the most intense waves that are responsible for the diffusion. Indeed, we found that the role of the 80 $\%$ less intense whistler waves is equivalent to that of the 20 $\%$ most intense ones. 
It should be noted that these results are coherent with observations by \citet{kajdic_suprathermal_2016}. Indeed, using Cluster data at 1 AU, they showed that the Strahl PAW is between 2$^\circ$ and 12$^\circ$ larger during intervals when whistlers are present than during intervals when they are absent. The whistlers observed were also predominantly quasi-aligned. 
The diffusion of halo electrons by Strahl-aligned whistler waves is very efficient. Specifically, quasi-aligned whistlers ($\theta \in [0,15] ^\circ$) make a very effective contribution to the isotropization of the halo electrons. \\

\begin{figure*}
   \centering
   \includegraphics[width=\linewidth]{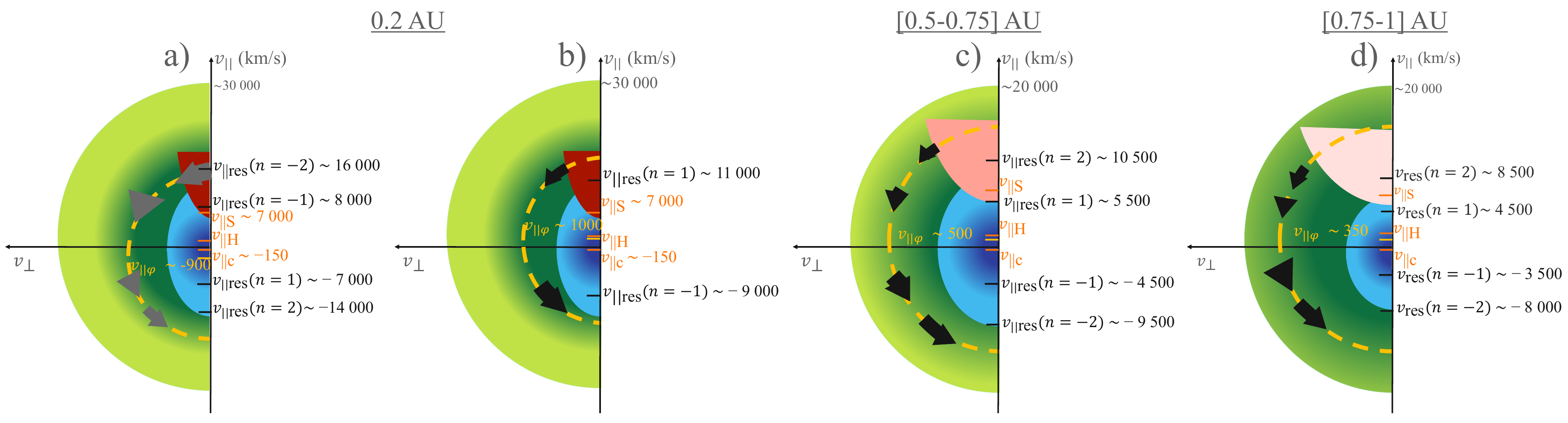}
      \caption{Scheme of the evolution of the e-VDF in the ($v_{\perp}$, $v_{||}$) plane and its interaction with whistler waves as a function of heliocentric distance. \textit{Panel a)} represents the effect of counter-streaming whistler waves at 0.2 AU. \textit{Panels b),c),d)} show the effect of Strahl-aligned whistler waves between 0.2 and 1 AU. The blue circles represent the core electrons, the green circles the halo electrons, and the red/pink beams the Strahl electrons. The intensity of the color represents the population's relative density. The resonance velocities ($v_{\rm res||}$) for $n=\pm 1$ and $n=\pm 2$, the drift velocities of the core, halo and Strahl (noted $v_{\rm c||}$,$v_{\rm H||}$,$v_{\rm S||}$, respectively), and the parallel phase velocities $v_{\varphi||}$ of the waves are indicated along the $v_{||}$ axis. The resonance velocities with $n=\pm 3$ are not represented for clarity. 
      One of the single-wave characteristics (centered around $v_{\varphi||}$) is represented by an orange dotted line. The path of the diffusing electrons is indicated by the arrows. This diffusion takes place around parallel resonance speeds.
      The role of the diffusing electrons in the amplification (resp., damping) of the waves is represented by black (resp., grey) arrows. 
        }
      \label{schema_diffusion}
   \end{figure*}

Figure \ref{schema_diffusion} shows these results together with schematics of the electron velocity distribution functions (e-VDFs) for different heliocentric distances, with the aim of representing how whistler waves interact with the distribution. 
The parallel $||$ and perpendicular $\perp$ temperatures of the core electrons observed by \cite{stverak_electron_2015} are used to represent the anisotropy of this population. The halo is assumed to be isotropic. The energy distribution of the different populations and their velocity drifts in the proton reference frame are taken from \cite{halekas_electrons_2020,halekas_radial_2022} for Figures \ref{schema_diffusion}a and \ref{schema_diffusion}b, \cite{bercic_whistler_2021} for Figure \ref{schema_diffusion}c and \cite{tong_whistler_2019} for Figure \ref{schema_diffusion}d. 
The phase and resonance velocities are the mean values observed in this study, and we can note that they were found to decrease with distance. As explained in Section \ref{Statistical analysis}, to calculate the wave vector, we use the cold plasma dispersion equation \cite{lyons_pitch_1974}, which allows us to calculate the phase velocity and the resonance velocity (Equation \ref{resonance}). 
 The electron diffusion occurs along circles centered on the parallel wave phase velocity (orange dotted lines, called single-wave characteristics), and the net flux of particles is directed towards a less dense area of the phase space. If the electrons gain (resp., lose) energy in the plasma frame, then they damp (resp., amplify) the waves \citep{kennel_resonant_1967,lyons_pitch_1974,verscharen_electron-driven_2022}. 

Each panel of Figure \ref{schema_diffusion} illustrates a diffusion process caused by whistler waves that can be described as follows. First, an instability is triggered by the free energy available in the distribution function. Then, the instability creates a wave that interacts with the electrons of the distribution function and diffuses them in the phase space. The objective of Figure \ref{schema_diffusion} is not to represent the instabilities that can create the waves but to schematically represent the evolution of the distribution function and the role of the whistler waves in this evolution. Nevertheless, at 0.2 AU, for counter-streaming whistlers, the instability can be caused by trapped electrons as proposed by \cite{karbashewski_whistler_2023}. For Strahl-aligned whistlers, the instabilities can be the WHFI \citep{gary_electron_1975,tong_whistler_2019}, the temperature anisotropy \citep{sagdeev_instability_1960,kennel_limit_1966,jagarlamudi_whistler_2020,vasko_quasi-parallel_2020} or the sunward suprathermal electron deficit \citep{bercic_whistler_2021}. 
 \\


At 0.2 AU we represent the distribution function with a high density ratio between the Strahl and the halo and with a strongly focused Strahl, as it is well known \citep{maksimovic_radial_2005,stverak_radial_2009,bercic_scattering_2019}. The diffusion of Strahl electrons by the counter-streaming whistlers is very efficient (Figure \ref{schema_diffusion}a). However, the diffusion of Strahl electrons by Strahl-aligned whistlers is less efficient (Figure \ref{schema_diffusion}b). It is likely that these diffusion processes at 0.2 AU explain the increase of the Strahl PAW as well as of the relative proportion of halo electrons, that we schematize in Figure \ref{schema_diffusion}c).

An increase in relative halo density and Strahl PAW is then represented (Figures \ref{schema_diffusion}c to \ref{schema_diffusion}d) \citep{maksimovic_radial_2005,stverak_radial_2009,bercic_scattering_2019}. The results of previous sections suggest that whistler waves verifying $\theta \in [15,45]^\circ$ explain the increase in the Strahl PAW and probably the increase in the relative density of the halo. The observed evolution of the wave occurence (Figure \ref{occurence}) can be interpreted as follows. The growth of the relative halo density could explain the increase in the occurrence of the waves in the observations between 0.3 and 0.6 AU. We can then suppose that the quasi-aligned waves saturate the instabilities, which could then explain the decrease in the occurrence between 0.6 and 1 AU.
For Figures \ref{schema_diffusion}b to \ref{schema_diffusion}d the $n<0$ resonances are very efficient and participate in the halo istropization whereas $n>0$ resonances participate slowly to the Strahl diffusion.

To conclude: 
\begin{itemize}
    \item We perform the largest statistical study of whistler waves in the solar wind to date (about 110,000 wave packets). This statistic contains all important whistler parameters between 0.2 and 1 AU to understand the wave-particle interactions. In particular, we characterize whistler wave occurrence, amplitude, propagation angle, and radial direction of propagation. 

   \item Between 0.3 and 1 AU, we observe an overwhelming majority of whistlers propagating in the Strahl direction (Strahl-aligned) and nearly aligned with the magnetic field. The few whistler waves found to propagate sunward are within switchbacks with a change of sign of the radial component, so that they are still aligned with the Strahl. 
   At 0.2 AU, we observe both Strahl-aligned and counter-streaming (propagating in the opposite direction to the Strahl) whistler waves. 

  \item Between 0.3 and 1 AU, whistlers propagating at an angle verifying $\theta \in [15,45]^\circ$ (the origin of which has not
  yet been fully determined) are sufficiently efficient (of the order of 1 $^\circ/\si{\second}$) to explain the observed increase of the Strahl PAW in spite of the fact that they represent only 7$\%$ of the cases. These waves probably also explain the observed transfer from the Strahl to the halo.
  At 0.2 AU the counter-streaming whistler waves are more efficient in diffusing the Strahl than the Strahl-aligned waves by two orders of magnitude. However, we had to restrict our analysis to the first perihelion of PSP and the statistics is small. We have developed a method to retrieve whistler waves properties for other perihelia \citep{colomban_reconstruction_2023} which will enable us to increase the statistics at 0.2 AU and below and to better characterize whistler occurrences and generation mechanisms.
\end{itemize}

\begin{acknowledgements}
L.C, M.K, V.K, C.F, M.M, and M.B acknowledge CNES, which supported this work and the development of RPW and SWA/EAS on Solar Orbiter and FIELDS/SCM on Parker Solar Probe. The UK Space Agency (grant ST/T001062/1) supports the Solar Orbiter's magnetometer operations. The Solar Orbiter SWA research at UCL/MSSL is currently supported by the STFC grants ST/T001356/1 and ST/S000240/1. Johns Hopkins Applied Physics Laboratory designed, built, and operates, Parker Solar Probe (contract NNN06AA01C).  Parker Solar Probe is part of NASA’s Living with a Star (LWS) program.  We thank the FIELDS and the RPW teams for providing the data. The data used in this work can be found in the Solar Orbiter archive (http://soar.esac.esa.int/soar/) and NASA CDAWeb (https://cdaweb.gsfc.nasa.gov/index.html).
\end{acknowledgements}

\bibliographystyle{aa}
\bibliography{papier_diffusion2}

\begin{thebibliography}{129}
\expandafter\ifx\csname natexlab\endcsname\relax\def\natexlab#1{#1}\fi

\bibitem[{Agapitov {et~al.}(2013)Agapitov, Artemyev, Krasnoselskikh, Khotyaintsev, Mourenas, Breuillard, Balikhin, \& Rolland}]{agapitov_statistics_2013}
Agapitov, O., Artemyev, A., Krasnoselskikh, V., {et~al.} 2013, Journal of Geophysical Research: Space Physics, 118, 3407

\bibitem[{Agapitov {et~al.}(2022)Agapitov, Drake, Swisdak, Bale, Horbury, Kasper, MacDowall, Mozer, Phan, Pulupa, Raouafi, \& Velli}]{agapitov_flux_2022}
Agapitov, O.~V., Drake, J.~F., Swisdak, M., {et~al.} 2022, The Astrophysical Journal, 925, 213

\bibitem[{Agapitov {et~al.}(2020)Agapitov, Dudok~de Wit, Mozer, Bonnell, Drake, Malaspina, Krasnoselskikh, Bale, Whittlesey, Case, Chaston, Froment, Goetz, Goodrich, Harvey, Kasper, Korreck, Larson, Livi, MacDowall, Pulupa, Revillet, Stevens, \& Wygant}]{agapitov_sunward-propagating_2020}
Agapitov, O.~V., Dudok~de Wit, T., Mozer, F.~S., {et~al.} 2020, The Astrophysical Journal, 891, L20

\bibitem[{Artemyev {et~al.}(2012)Artemyev, Agapitov, Krasnoselskikh, Breuillard, \& Rolland}]{artemyev_statistical_2012}
Artemyev, A., Agapitov, O., Krasnoselskikh, V., Breuillard, H., \& Rolland, G. 2012, Journal of Geophysical Research: Space Physics, 117

\bibitem[{Bale {et~al.}(2019)Bale, Badman, Bonnell, Bowen, Burgess, Case, Cattell, Chandran, Chaston, Chen, Drake, de~Wit, Eastwood, Ergun, Farrell, Fong, Goetz, Goldstein, Goodrich, Harvey, Horbury, Howes, Kasper, Kellogg, Klimchuk, Korreck, Krasnoselskikh, Krucker, Laker, Larson, MacDowall, Maksimovic, Malaspina, Martinez~Oliveros, McComas, Meyer-Vernet, Moncuquet, Mozer, Phan, Pulupa, Raouafi, Salem, Stansby, Stevens, Szabo, Velli, Woolley, \& Wygant}]{bale_highly_2019}
Bale, S.~D., Badman, S.~T., Bonnell, J.~W., {et~al.} 2019, Nature, 576, 237, number: 7786 Publisher: Nature Publishing Group

\bibitem[{Bale {et~al.}(2016)Bale, Goetz, Harvey, Turin, Bonnell, Dudok¬†de¬†Wit, Ergun, MacDowall, Pulupa, Andre, Bolton, Bougeret, Bowen, Burgess, Cattell, Chandran, Chaston, Chen, Choi, Connerney, Cranmer, Diaz-Aguado, Donakowski, Drake, Farrell, Fergeau, Fermin, Fischer, Fox, Glaser, Goldstein, Gordon, Hanson, Harris, Hayes, Hinze, Hollweg, Horbury, Howard, Hoxie, Jannet, Karlsson, Kasper, Kellogg, Kien, Klimchuk, Krasnoselskikh, Krucker, Lynch, Maksimovic, Malaspina, Marker, Martin, MartinezOliveros, McCauley, McComas, McDonald, Meyer~Vernet, Moncuquet, Monson, Mozer, Murphy, Odom, Oliverson, Olson, Parker, Pankow, Phan, Quataert, Quinn, Ruplin, Salem, Seitz, Sheppard, Siy, Stevens, Summers, Szabo, Timofeeva, Vaivads, Velli, Yehle, Werthimer, \& Wygant}]{bale_fields_2016}
Bale, S.~D., Goetz, K., Harvey, P.~R., {et~al.} 2016, Space Science Reviews, 204, 49

\bibitem[{Bale {et~al.}(2013)Bale, Pulupa, Salem, Chen, \& Quataert}]{bale_electron_2013}
Bale, S.~D., Pulupa, M., Salem, C., Chen, C. H.~K., \& Quataert, E. 2013, The Astrophysical Journal, 769, L22, arXiv:1303.0932 [astro-ph, physics:physics]

\bibitem[{Bercic {et~al.}(2019)Bercic, Maksimovic, Landi, \& Matteini}]{bercic_scattering_2019}
Bercic, L., Maksimovic, M., Landi, S., \& Matteini, L. 2019, Monthly Notices of the Royal Astronomical Society, 486, 3404, arXiv:1904.08272 [astro-ph, physics:physics]

\bibitem[{Bercic {et~al.}(2021{\natexlab{a}})Bercic, Landi, \& Maksimovic}]{bercic_interplay_2021}
Bercic, L., Landi, S., \& Maksimovic, M. 2021{\natexlab{a}}, Journal of Geophysical Research: Space Physics, 126, e2020JA028864

\bibitem[{Bercic {et~al.}(2020)Bercic, Larson, Whittlesey, Maksimovic, Badman, Landi, Matteini, Bale, Bonnell, Case, Dudok~de Wit, Goetz, Harvey, Kasper, Korreck, Livi, MacDowall, Malaspina, Pulupa, \& Stevens}]{bercic_coronal_2020}
Bercic, L., Larson, D., Whittlesey, P., {et~al.} 2020, The Astrophysical Journal, 892, 88

\bibitem[{Bercic {et~al.}(2021{\natexlab{b}})Bercic, Verscharen, Owen, Colomban, Kretzschmar, Chust, Maksimovic, Kataria, Behar, Berthomier, Bruno, Fortunato, Kelly, Khotyaintsev, Lewis, Livi, Louarn, Mele, Nicolaou, Watson, \& Wicks}]{bercic_whistler_2021}
Bercic, L., Verscharen, D., Owen, C.~J., {et~al.} 2021{\natexlab{b}}, Astronomy \& Astrophysics, 656, A31, arXiv:2107.10645 [astro-ph, physics:physics]

\bibitem[{Boldyrev \& Horaites(2019)}]{boldyrev_kinetic_2019}
Boldyrev, S. \& Horaites, K. 2019, Monthly Notices of the Royal Astronomical Society, 489, 3412

\bibitem[{Case {et~al.}(2020)Case, Kasper, Stevens, Korreck, Paulson, Daigneau, Caldwell, Freeman, Henry, Klingensmith, Bookbinder, Robinson, Berg, Tiu, Wright, Reinhart, Curtis, Ludlam, Larson, Whittlesey, Livi, Klein, \& Martinovic}]{case_solar_2020}
Case, A.~W., Kasper, J.~C., Stevens, M.~L., {et~al.} 2020, The Astrophysical Journal Supplement Series, 246, 43

\bibitem[{Cattell {et~al.}(2022)Cattell, Breneman, Dombeck, Hanson, Johnson, Halekas, Bale, Wit, Goetz, Goodrich, Malaspina, Pulupa, Case, Kasper, Larson, Stevens, \& Whittlesey}]{cattell_parker_2022}
Cattell, C., Breneman, A., Dombeck, J., {et~al.} 2022, The Astrophysical Journal Letters, 924, L33, publisher: The American Astronomical Society

\bibitem[{Cattell {et~al.}(2021{\natexlab{a}})Cattell, Breneman, Dombeck, Short, Wygant, Halekas, Case, Kasper, Larson, Stevens, Whittesley, Bale, Dudok~de Wit, Goodrich, MacDowall, Moncuquet, Malaspina, \& Pulupa}]{cattell_parker_2021}
Cattell, C., Breneman, A., Dombeck, J., {et~al.} 2021{\natexlab{a}}, The Astrophysical Journal Letters, 911, L29

\bibitem[{Cattell {et~al.}(2021{\natexlab{b}})Cattell, Short, Breneman, Halekas, Whittesley, Larson, Kasper, Stevens, Case, Moncuquet, Bale, Bonnell, Dudok~de Wit, Goetz, Harvey, MacDowall, Malaspina, Maksimovic, Pulupa, \& Goodrich}]{cattell_narrowband_2021}
Cattell, C., Short, B., Breneman, A., {et~al.} 2021{\natexlab{b}}, Astronomy \& Astrophysics, 650, A8

\bibitem[{Cattell {et~al.}(2020)Cattell, Short, Breneman, \& Grul}]{cattell_narrowband_2020}
Cattell, C., Short, B., Breneman, A.~W., \& Grul, P. 2020, The Astrophysical Journal, 897, 126

\bibitem[{Cattell \& Vo(2021)}]{cattell_modeling_2021}
Cattell, C. \& Vo, T. 2021, The Astrophysical Journal Letters, 914, L33

\bibitem[{Chust {et~al.}(2021)Chust, Kretzschmar, Graham, Le~Contel, Retino, Alexandrova, Berthomier, Hadid, Sahraoui, Jeandet, Leroy, Pellion, Bouzid, Katra, Piberne, Khotyaintsev, Vaivads, Krasnoselskikh, Soucek, Santolik, Lorfevre, Plettemeier, Steller, Steverak, Travnicek, Vecchio, Maksimovic, Bale, Horbury, OBrien, Evans, \& Angelini}]{chust_observations_2021}
Chust, T., Kretzschmar, M., Graham, D.~B., {et~al.} 2021, Astronomy \& Astrophysics, 656, A17

\bibitem[{Colomban {et~al.}(2023)Colomban, Agapitov, Krasnoselskikh, Kretzschmar, Dudok~de Wit, Karbashewski, Mozer, Bonnell, Bale, Malaspina, \& Raouafi}]{colomban_reconstruction_2023}
Colomban, L., Agapitov, O.~V., Krasnoselskikh, V., {et~al.} 2023, Journal of Geophysical Research: Space Physics, 128, e2023JA031427, \_eprint: https://onlinelibrary.wiley.com/doi/pdf/10.1029/2023JA031427

\bibitem[{Dudok De~Wit {et~al.}(2020)Dudok De~Wit, Krasnoselskikh, Bale, Bonnell, Bowen, Chen, Froment, Goetz, Harvey, Jagarlamudi, Larosa, MacDowall, Malaspina, Matthaeus, Pulupa, Velli, \& Whittlesey}]{dudok_de_wit_switchbacks_2020}
Dudok De~Wit, T., Krasnoselskikh, V., Bale, S.~D., {et~al.} 2020, The Astrophysical Journal Supplement Series, 246, 39

\bibitem[{Dudok De~Wit {et~al.}(2022)Dudok De~Wit, Krasnoselskikh, Agapitov, Froment, Larosa, Bale, Bowen, Goetz, Harvey, Jannet, Kretzschmar, MacDowall, Malaspina, Martin, Page, Pulupa, \& Revillet}]{dudok_de_wit_first_2022}
Dudok De~Wit, T., Krasnoselskikh, V.~V., Agapitov, O., {et~al.} 2022, Journal of Geophysical Research: Space Physics, 127, e2021JA030018

\bibitem[{Feldman {et~al.}(1976)Feldman, Asbridge, Bame, Gary, Montgomery, \& Zink}]{feldman_evidence_1976}
Feldman, W.~C., Asbridge, J.~R., Bame, S.~J., {et~al.} 1976, Journal of Geophysical Research (1896-1977), 81, 5207, \_eprint: https://onlinelibrary.wiley.com/doi/pdf/10.1029/JA081i028p05207

\bibitem[{Feldman {et~al.}(1978)Feldman, Asbridge, Bame, Gosling, \& Lemons}]{feldman_characteristic_1978}
Feldman, W.~C., Asbridge, J.~R., Bame, S.~J., Gosling, J.~T., \& Lemons, D.~S. 1978, Journal of Geophysical Research: Space Physics, 83, 5285

\bibitem[{Feldman {et~al.}(1975)Feldman, Asbridge, Bame, Montgomery, \& Gary}]{feldman_solar_1975}
Feldman, W.~C., Asbridge, J.~R., Bame, S.~J., Montgomery, M.~D., \& Gary, S.~P. 1975, Journal of Geophysical Research, 80, 4181

\bibitem[{Fitzenreiter {et~al.}(1998)Fitzenreiter, Ogilvie, Chornay, \& Keller}]{fitzenreiter_observations_1998}
Fitzenreiter, R.~J., Ogilvie, K.~W., Chornay, D.~J., \& Keller, J. 1998, Geophysical Research Letters, 25, 249

\bibitem[{Fox {et~al.}(2016)Fox, Velli, Bale, Decker, Driesman, Howard, Kasper, Kinnison, Kusterer, Lario, Lockwood, McComas, Raouafi, \& Szabo}]{fox_solar_2016}
Fox, N.~J., Velli, M.~C., Bale, S.~D., {et~al.} 2016, Space Science Reviews, 204, 7

\bibitem[{Froment {et~al.}(2023)Froment, Agapitov, Krasnoselskikh, Karbashewski, Dudok De~Wit, Larosa, Colomban, Malaspina, Kretzschmar, Jagarlamudi, Bale, Bonnell, Mozer, \& Pulupa}]{froment_whistler_2023}
Froment, C., Agapitov, O.~V., Krasnoselskikh, V., {et~al.} 2023, Astronomy \& Astrophysics, 672, A135

\bibitem[{Gary \& Feldman(1977)}]{gary_solar_1977}
Gary, S.~P. \& Feldman, W.~C. 1977, Journal of Geophysical Research, 82, 1087

\bibitem[{Gary {et~al.}(1975)Gary, Feldman, Forslund, \& Montgomery}]{gary_electron_1975}
Gary, S.~P., Feldman, W.~C., Forslund, D.~W., \& Montgomery, M.~D. 1975, Geophysical Research Letters, 2, 79

\bibitem[{Gary {et~al.}(1999)Gary, Skoug, \& Daughton}]{gary_electron_1999}
Gary, S.~P., Skoug, R.~M., \& Daughton, W. 1999, Physics of Plasmas, 6, 2607

\bibitem[{Gary \& Wang(1996)}]{gary_whistler_1996}
Gary, S.~P. \& Wang, J. 1996, Journal of Geophysical Research: Space Physics, 101, 10749, \_eprint: https://onlinelibrary.wiley.com/doi/pdf/10.1029/96JA00323

\bibitem[{Glauert \& Horne(2005)}]{glauert_calculation_2005}
Glauert, S.~A. \& Horne, R.~B. 2005, Journal of Geophysical Research: Space Physics, 110

\bibitem[{Graham {et~al.}(2018)Graham, Rae, Owen, \& Walsh}]{graham_investigating_2018}
Graham, G.~A., Rae, I.~J., Owen, C.~J., \& Walsh, A.~P. 2018, The Astrophysical Journal, 855, 40

\bibitem[{Graham {et~al.}(2017)Graham, Rae, Owen, Walsh, Arridge, Gilbert, Lewis, Jones, Forsyth, Coates, \& Waite}]{graham_evolution_2017}
Graham, G.~A., Rae, I.~J., Owen, C.~J., {et~al.} 2017, Journal of Geophysical Research: Space Physics, 122, 3858

\bibitem[{Gurgiolo \& Goldstein(2017)}]{gurgiolo_absence_2017}
Gurgiolo, C. \& Goldstein, M.~L. 2017, Annales Geophysicae, 35, 71, publisher: Copernicus GmbH

\bibitem[{Gurnett \& Anderson(1977)}]{gurnett_plasma_1977}
Gurnett, D.~A. \& Anderson, R.~R. 1977, Journal of Geophysical Research (1896-1977), 82, 632, \_eprint: https://onlinelibrary.wiley.com/doi/pdf/10.1029/JA082i004p00632

\bibitem[{Halekas {et~al.}(2021{\natexlab{a}})Halekas, Bercic, Whittlesey, Larson, Livi, Berthomier, Kasper, Case, Stevens, Bale, MacDowall, \& Pulupa}]{halekas_sunward_2021}
Halekas, J.~S., Bercic, L., Whittlesey, P., {et~al.} 2021{\natexlab{a}}, The Astrophysical Journal, 916, 16

\bibitem[{Halekas {et~al.}(2022)Halekas, Whittlesey, Larson, Maksimovic, Livi, Berthomier, Kasper, Case, Stevens, Bale, MacDowall, \& Pulupa}]{halekas_radial_2022}
Halekas, J.~S., Whittlesey, P., Larson, D.~E., {et~al.} 2022, The Astrophysical Journal, 936, 53

\bibitem[{Halekas {et~al.}(2020)Halekas, Whittlesey, Larson, McGinnis, Maksimovic, Berthomier, Kasper, Case, Korreck, Stevens, Klein, Bale, MacDowall, Pulupa, Malaspina, Goetz, \& Harvey}]{halekas_electrons_2020}
Halekas, J.~S., Whittlesey, P., Larson, D.~E., {et~al.} 2020, The Astrophysical Journal Supplement Series, 246, 22

\bibitem[{Halekas {et~al.}(2021{\natexlab{b}})Halekas, Whittlesey, Larson, McGinnis, Bale, Berthomier, Case, Chandran, Kasper, Klein, Korreck, Livi, MacDowall, Maksimovic, Malaspina, Matteini, Pulupa, \& Stevens}]{halekas_electron_2021}
Halekas, J.~S., Whittlesey, P.~L., Larson, D.~E., {et~al.} 2021{\natexlab{b}}, Astronomy \& Astrophysics, 650, A15, arXiv:2010.10302 [astro-ph, physics:physics]

\bibitem[{Hammond {et~al.}(1996)Hammond, Feldman, McComas, Phillips, \& Forsyth}]{hammond_variation_1996}
Hammond, C.~M., Feldman, W.~C., McComas, D.~J., Phillips, J.~L., \& Forsyth, R.~J. 1996, Astronomy and Astrophysics, 316, 350

\bibitem[{Horbury {et~al.}(2020)Horbury, OBrien, Carrasco~Blazquez, Bendyk, Brown, Hudson, Evans, Oddy, Carr, Beek, Cupido, Bhattacharya, Dominguez, Matthews, Myklebust, Whiteside, Bale, Baumjohann, Burgess, Carbone, Cargill, Eastwood, Fletcher, Forsyth, Giacalone, Glassmeier, Goldstein, Hoeksema, Lockwood, Magnes, Maksimovic, Marsch, Matthaeus, Murphy, Nakariakov, Owen, Owens, Rodriguez-Pacheco, Richter, Riley, Russell, Schwartz, Vainio, Velli, Vennerstrom, Walsh, Wimmer-Schweingruber, Zank, Miller, Zouganelis, \& Walsh}]{horbury_solar_2020}
Horbury, T.~S., OBrien, H., Carrasco~Blazquez, I., {et~al.} 2020, Astronomy \& Astrophysics, 642, A9

\bibitem[{Horne(2003)}]{horne_resonant_2003}
Horne, R.~B. 2003, Geophysical Research Letters, 30, 1493

\bibitem[{Jagarlamudi {et~al.}(2020)Jagarlamudi, Alexandrova, Bercic, Dudok~de Wit, Krasnoselskikh, Maksimovic, \& Stverak}]{jagarlamudi_whistler_2020}
Jagarlamudi, V.~K., Alexandrova, O., Bercic, L., {et~al.} 2020, The Astrophysical Journal, 897, 118, arXiv: 2008.02334

\bibitem[{Jagarlamudi {et~al.}(2021)Jagarlamudi, Dudok~de Wit, Froment, Krasnoselskikh, Larosa, Bercic, Agapitov, Halekas, Kretzschmar, Malaspina, Moncuquet, Bale, Case, Kasper, Korreck, Larson, Pulupa, Stevens, \& Whittlesey}]{jagarlamudi_whistler_2021}
Jagarlamudi, V.~K., Dudok~de Wit, T., Froment, C., {et~al.} 2021, Astronomy \& Astrophysics, 650, A9

\bibitem[{Jannet {et~al.}(2021)Jannet, Dudok~de Wit, Krasnoselskikh, Kretzschmar, Fergeau, Bergerard~Timofeeva, Agrapart, Brochot, Chalumeau, Martin, Revillet, Bale, Maksimovic, Bowen, Brysbaert, Goetz, Guilhem, Harvey, Leray, \& Lorfevre}]{jannet_measurement_2021}
Jannet, G., Dudok~de Wit, T., Krasnoselskikh, V., {et~al.} 2021, Journal of Geophysical Research: Space Physics, 126

\bibitem[{Jeong {et~al.}(2022)Jeong, Abraham, Verscharen, Bercic, Stansby, Nicolaou, Owen, Wicks, Fazakerley, Agudelo~Rueda, \& Bakrania}]{jeong_stability_2022}
Jeong, S.~Y., Abraham, J.~B., Verscharen, D., {et~al.} 2022, The Astrophysical Journal Letters, 926, L26

\bibitem[{Kadomtsev \& Pogutse(1968)}]{kadomtsev_electric_1968}
Kadomtsev, B.~B. \& Pogutse, O.~P. 1968, Soviet Journal of Experimental and Theoretical Physics, 26, 1146, aDS Bibcode: 1968JETP...26.1146K

\bibitem[{Kadijc {et~al.}(2016)Kajdic, Alexandrova, Maksimovic, Lacombe, \& Fazakerley}]{kajdic_suprathermal_2016}
Kajdic, P., Alexandrova, O., Maksimovic, M., Lacombe, C., \& Fazakerley, A.~N. 2016, The Astrophysical Journal, 833, 172

\bibitem[{Karbashewski {et~al.}(2023)Karbashewski, Agapitov, Kim, Mozer, Bonnell, Froment, Dudok~de Wit, Bale, Malaspina, \& Raouafi}]{karbashewski_whistler_2023}
Karbashewski, S., Agapitov, O., Kim, H., {et~al.} 2023, The Astrophysical Journal, 947, 73

\bibitem[{Karpman {et~al.}(1975)Karpman, Istomin, \& Shklyar}]{karpman_effects_1975}
Karpman, V.~I., Istomin, J.~N., \& Shklyar, D.~R. 1975, Physica Scripta, 11, 278

\bibitem[{Kasper {et~al.}(2016)Kasper, Abiad, Austin, Balat-Pichelin, Bale, Belcher, Berg, Bergner, Berthomier, Bookbinder, Brodu, Caldwell, Case, Chandran, Cheimets, Cirtain, Cranmer, Curtis, Daigneau, Dalton, Dasgupta, DeTomaso, Diaz-Aguado, Djordjevic, Donaskowski, Effinger, Florinski, Fox, Freeman, Gallagher, Gary, Gauron, Gates, Goldstein, Golub, Gordon, Gurnee, Guth, Halekas, Hatch, Heerikuisen, Ho, Hu, Johnson, Jordan, Korreck, Larson, Lazarus, Li, Livi, Ludlam, Maksimovic, McFadden, Marchant, Maruca, McComas, Messina, Mercer, Park, Peddie, Pogorelov, Reinhart, Richardson, Robinson, Rosen, Skoug, Slagle, Steinberg, Stevens, Szabo, Taylor, Tiu, Turin, Velli, Webb, Whittlesey, Wright, Wu, \& Zank}]{kasper_solar_2016}
Kasper, J.~C., Abiad, R., Austin, G., {et~al.} 2016, Space Science Reviews, 204, 131

\bibitem[{Kennel \& Engelmann(1966)}]{kennel_velocity_1966}
Kennel, C.~F. \& Engelmann, F. 1966, The Physics of Fluids, 9, 2377, publisher: American Institute of PhysicsAIP

\bibitem[{Kennel \& Petschek(1966)}]{kennel_limit_1966}
Kennel, C.~F. \& Petschek, H.~E. 1966, Journal of Geophysical Research (1896-1977), 71, 1

\bibitem[{Kennel \& Wong(1967)}]{kennel_resonant_1967}
Kennel, C.~F. \& Wong, H.~V. 1967, Journal of Plasma Physics, 1, 75

\bibitem[{Khotyaintsev {et~al.}(2021)Khotyaintsev, Graham, Vaivads, Steinvall, Edberg, Eriksson, Johansson, Sorriso-Valvo, Maksimovic, Bale, Chust, Krasnoselskikh, Kretzschmar, Lorfevre, Plettemeier, Soucek, Steller, Steverak, Travnicek, Vecchio, Horbury, Evans, \& Angelini}]{khotyaintsev_density_2021}
Khotyaintsev, Y.~V., Graham, D.~B., Vaivads, A., {et~al.} 2021, Astronomy \& Astrophysics, 656, A19

\bibitem[{Komarov {et~al.}(2018)Komarov, Schekochihin, Churazov, \& Spitkovsky}]{komarov_self-inhibiting_2018}
Komarov, S., Schekochihin, A.~A., Churazov, E., \& Spitkovsky, A. 2018, Journal of Plasma Physics, 84, 905840305

\bibitem[{Krafft \& Volokitin(2003)}]{krafft_interaction_2003}
Krafft, C. \& Volokitin, A. 2003, Annales Geophysicae, 21, 1393, publisher: Copernicus GmbH


\bibitem[{Krafft \& Volokitin(2003)}]{krafft_interaction_2003}
Krafft, C. \& Volokitin, A. 2003, Annales Geophysicae, 21, 1393, publisher: Copernicus GmbH

\bibitem[{Krasnoselskikh {et~al.}(2020)Krasnoselskikh, Larosa, Agapitov, Dudok~de Wit, Moncuquet, Mozer, Stevens, Bale, Bonnell, Froment, Goetz, Goodrich, Harvey, Kasper, MacDowall, Malaspina, Pulupa, Raouafi, Revillet, Velli, \& Wygant}]{krasnoselskikh_localized_2020}
Krasnoselskikh, V., Larosa, A., Agapitov, O., {et~al.} 2020, The Astrophysical Journal, 893, 93

\bibitem[{Kretzschmar {et~al.}(2021)Kretzschmar, Chust, Krasnoselskikh, Graham, Colomban, Maksimovic, Khotyaintsev, Soucek, Steinvall, Santolik, Jannet, Brochot, Le~Contel, Vecchio, Bonnin, Bale, Froment, Larosa, Bergerard-Timofeeva, Fergeau, Lorfevre, Plettemeier, Steller, Steverak, Travnicek, Vaivads, Horbury, Evans, Angelini, Owen, \& Louarn}]{kretzschmar_whistler_2021}
Kretzschmar, M., Chust, T., Krasnoselskikh, V., {et~al.} 2021, Astronomy \& Astrophysics, 656, A24

\bibitem[{Kuzichev {et~al.}(2019)Kuzichev, Vasko, Soto-Chavez, Tong, Artemyev, Bale, \& Spitkovsky}]{kuzichev_nonlinear_2019}
Kuzichev, I.~V., Vasko, I.~Y., Soto-Chavez, A.~R., {et~al.} 2019, The Astrophysical Journal, 882, 81, arXiv:1907.04878 [physics]

\bibitem[{Lacombe {et~al.}(2014)Lacombe, Alexandrova, Matteini, Santolik, Cornilleau-Wehrlin, Mangeney, de~Conchy, \& Maksimovic}]{lacombe_whistler_2014}
Lacombe, C., Alexandrova, O., Matteini, L., {et~al.} 2014, The Astrophysical Journal, 796, 5

\bibitem[{Lazar {et~al.}(2019)Lazar, Lopez, Shaaban, Poedts, \& Fichtner}]{lazar_whistler_2019}
Lazar, M., Lopez, R.~A., Shaaban, S.~M., Poedts, S., \& Fichtner, H. 2019, Astrophysics and Space Science, 364, 171

\bibitem[{Lazar {et~al.}(2020)Lazar, Pierrard, Poedts, \& Fichtner}]{lazar_characteristics_2020}
Lazar, M., Pierrard, V., Poedts, S., \& Fichtner, H. 2020, Astronomy \& Astrophysics, 642, A130

\bibitem[{Lazar {et~al.}(2013)Lazar, Poedts, \& Michno}]{lazar_electromagnetic_2013}
Lazar, M., Poedts, S., \& Michno, M.~J. 2013, Astronomy \& Astrophysics, 554, A64

\bibitem[{Lazar {et~al.}(2011)Lazar, Poedts, \& Schlickeiser}]{lazar_instability_2011}
Lazar, M., Poedts, S., \& Schlickeiser, R. 2011, Monthly Notices of the Royal Astronomical Society, 410, 663

\bibitem[{Lazar {et~al.}(2014)Lazar, Poedts, \& Schlickeiser}]{lazar_interplay_2014}
Lazar, M., Poedts, S., \& Schlickeiser, R. 2014, Journal of Geophysical Research: Space Physics, 119, 9395

\bibitem[{Lazar {et~al.}(2015)Lazar, Poedts, Schlickeiser, \& Dumitrache}]{lazar_towards_2015}
Lazar, M., Poedts, S., Schlickeiser, R., \& Dumitrache, C. 2015, Monthly Notices of the Royal Astronomical Society, 446, 3022

\bibitem[{Lazar {et~al.}(2018)Lazar, Yoon, Lopez, \& Moya}]{lazar_electromagnetic_2018}
Lazar, M., Yoon, P.~H., Lopez, R.~A., \& Moya, P.~S. 2018, Journal of Geophysical Research: Space Physics, 123, 6

\bibitem[{Lemaire \& Scherer(1970)}]{lemaire_model_1970}
Lemaire, J. \& Scherer, M. 1970, Planetary and Space Science, 18, 103

\bibitem[{Lyons(1974{\natexlab{a}})}]{lyons_general_1974}
Lyons, L.~R. 1974{\natexlab{a}}, Journal of Plasma Physics, 12, 45

\bibitem[{Lyons(1974{\natexlab{b}})}]{lyons_pitch_1974}
Lyons, L.~R. 1974{\natexlab{b}}, Journal of Plasma Physics, 12, 417

\bibitem[{Lyons {et~al.}(1971)Lyons, Thorne, \& Kennel}]{lyons_electron_1971}
Lyons, L.~R., Thorne, R.~M., \& Kennel, C.~F. 1971, Journal of Plasma Physics, 6, 589

\bibitem[{Lyons {et~al.}(1972)Lyons, Thorne, \& Kennel}]{lyons_pitch-angle_1972}
Lyons, L.~R., Thorne, R.~M., \& Kennel, C.~F. 1972, Journal of Geophysical Research, 77, 3455

\bibitem[{Lopez {et~al.}(2019)Lopez, Shaaban, Lazar, Poedts, Yoon, Micera, \& Lapenta}]{lopez_particle--cell_2019}
Lopez, R.~A., Shaaban, S.~M., Lazar, M., {et~al.} 2019, The Astrophysical Journal, 882, L8

\bibitem[{Macneil {et~al.}(2020)Macneil, Owens, Lockwood, Steverak, \& Owen}]{macneil_radial_2020}
Macneil, A.~R., Owens, M.~J., Lockwood, M., Steverak, S., \& Owen, C.~J. 2020, Solar Physics, 295, 16

\bibitem[{Maksimovic {et~al.}(2020)Maksimovic, Bale, Chust, Khotyaintsev, Krasnoselskikh, Kretzschmar, Plettemeier, Rucker, Soucek, Steller, Steverak, Travnicek, Vaivads, Chaintreuil, Dekkali, Alexandrova, Astier, Barbary, Bonnin, Boughedada, Cecconi, Chapron, Chariet, Collin, de~Conchy, Dias, Lamy, Leray, Lion, Malac-Allain, Matteini, Nguyen, Pantellini, Parisot, Plasson, Thijs, Vecchio, Fratter, Bellouard, Lorfevre, Danto, Julien, Guilhem, Fiachetti, Sanisidro, Laffaye, Gonzalez, Pontet, Jannet, Fergeau, Brochot, Cassam-Chenai, Dudok~de Wit, Timofeeva, Vincent, Agrapart, Delory, Turin, Jeandet, Leroy, Pellion, Bouzid, Katra, Piberne, Recart, Santolik, Bylander, Cripps, Cully, Eriksson, Jansson, Johansson, Karlsson, Puccio, Panchenko, Berthomier, Goetz, Hellinger, Horbury, Issautier, Kontar, Krucker, Le~Contel, Louarn, Martinoviƒá, Owen, Retino,
  Sahraoui, Wimmer-Schweingruber, Zaslavsky, \& Zouganelis}]{maksimovic_solar_2020}
Maksimovic, M., Bale, S.~D., Chust, T., {et~al.} 2020, Astronomy \& Astrophysics, 642, A12

\bibitem[{Maksimovic {et~al.}(1997)Maksimovic, Pierrard, \& Lemaire}]{maksimovic_kinetic_1997}
Maksimovic, M., Pierrard, V., \& Lemaire, J.~F. 1997, Astronomy and Astrophysics, 324, 725, aDS Bibcode: 1997A\&A...324..725M

\bibitem[{Maksimovic {et~al.}(2005)Maksimovic, Zouganelis, Chaufray, Issautier, Scime, Littleton, Marsch, McComas, Salem, Lin, \& Elliott}]{maksimovic_radial_2005}
Maksimovic, M., Zouganelis, I., Chaufray, J.-Y., {et~al.} 2005, Journal of Geophysical Research: Space Physics, 110

\bibitem[{Malaspina {et~al.}(2016)Malaspina, Ergun, Bolton, Kien, Summers, Stevens, Yehle, Karlsson, Hoxie, Bale, \& Goetz}]{malaspina_digital_2016}
Malaspina, D.~M., Ergun, R.~E., Bolton, M., {et~al.} 2016, Journal of Geophysical Research: Space Physics, 121, 5088

\bibitem[{Marsch(2006)}]{marsch_kinetic_2006}
Marsch, E. 2006, Living Reviews in Solar Physics, 3

\bibitem[{Means(1972)}]{means_use_1972}
Means, J.~D. 1972, Journal of Geophysical Research (1896-1977), 77, 5551, \_eprint: https://onlinelibrary.wiley.com/doi/pdf/10.1029/JA077i028p05551

\bibitem[{Micera {et~al.}(2021)Micera, Zhukov, Lopez, Boella, Tenerani, Velli, Lapenta, \& Innocenti}]{micera_role_2021}
Micera, A., Zhukov, A.~N., Lopez, R.~A., {et~al.} 2021, The Astrophysical Journal, 919, 42, arXiv:2106.15975 [astro-ph, physics:physics]

\bibitem[{Micera {et~al.}(2020)Micera, Zhukov, Lopez, Innocenti, Lazar, Boella, \& Lapenta}]{micera_particle--cell_2020}
Micera, A., Zhukov, A.~N., Lopez, R.~A., {et~al.} 2020, The Astrophysical Journal Letters, 903, L23, arXiv:2010.10832 [astro-ph]

\bibitem[{Mueller {et~al.}(2013)Mueller, Marsden, StCyr, \& Gilbert}]{mueller_solar_2013}
Mueller, D., Marsden, R.~G., StCyr, O.~C., \& Gilbert, H.~R. 2013, Solar Physics, 285, 25, arXiv:1207.4579 [astro-ph, physics:physics]

\bibitem[{Mueller {et~al.}(2020)Mueller, St.~Cyr, Zouganelis, Gilbert, Marsden, Nieves-Chinchilla, Antonucci, Auchere, Berghmans, Horbury, Howard, Krucker, Maksimovic, Owen, Rochus, Rodriguez-Pacheco, Romoli, Solanki, Bruno, Carlsson, Fludra, Harra, Hassler, Livi, Louarn, Peter, Schehle, Teriaca, del Toro~Iniesta, Wimmer-Schweingruber, Marsch, Velli, De~Groof, Walsh, \& Williams}]{muller_solar_2020}
Mueller, D., St.~Cyr, O.~C., Zouganelis, I., {et~al.} 2020, Astronomy \& Astrophysics, 642, A1

\bibitem[{Neubauer {et~al.}(1977)Neubauer, Musmann, \& Dehmel}]{neubauer_fast_1977}
Neubauer, F.~M., Musmann, G., \& Dehmel, G. 1977, Journal of Geophysical Research (1896-1977), 82, 3201

\bibitem[{Owen {et~al.}(2020)Owen, Bruno, Livi, Louarn, Al~Janabi, Allegrini, Amoros, Baruah, Barthe, Berthomier, Bordon, Brockley-Blatt, Brysbaert, Capuano, Collier, DeMarco, Fedorov, Ford, Fortunato, Fratter, Galvin, Hancock, Heirtzler, Kataria, Kistler, Lepri, Lewis, Loeffler, Marty, Mathon, Mayall, Mele, Ogasawara, Orlandi, Pacros, Penou, Persyn, Petiot, Phillips, Raines, Reden, Rouillard, Rousseau, Rubiella, Seran, Spencer, Thomas, Trevino, Verscharen, Wurz, Alapide, Amoruso, Anekallu, Arciuli, Arnett, Ascolese, Bancroft, Bland, Brysch, Calvanese, Castronuovo, Chornay, Clemens, Coker, Collinson, Dandouras, Darnley, Davies, Davison, De~Los~Santos, Devoto, Dirks, Edlund, Fazakerley, Ferris, Frost, Fruit, Garat, Gibson, Gilbert, de~Giosa, Gradone, Hailey, Horbury, Hunt, Jacquey, Johnson, Lavraud, Lawrenson, Leblanc, Lockhart, Maksimovic, Malpus, Marcucci, Mazelle, Monti, Myers, Nguyen, Rodriguez-Pacheco, Phillips, Popecki, Rees, Rogacki, Ruane, Rust, Salatti,
  Sauvaud, Stakhiv, Stange, Stubbs, Taylor, Techer, Terrier, Thibodeaux, Urdiales, Varsani, Walsh, Watson, Wheeler, Willis, Wimmer-Schweingruber, Winter, Yardley, \& Zouganelis}]{owen_solar_2020}
Owen, C.~J., Bruno, R., Livi, S., {et~al.} 2020, Astronomy \& Astrophysics, 642, A16

\bibitem[{Pagel {et~al.}(2007)Pagel, Gary, de~Koning, Skoug, \& Steinberg}]{pagel_scattering_2007}
Pagel, C., Gary, S.~P., de~Koning, C.~A., Skoug, R.~M., \& Steinberg, J.~T. 2007, Journal of Geophysical Research: Space Physics, 112

\bibitem[{Parail \& Pogutse(1978)}]{parail_kinetic_1978}
Parail, V.~V. \& Pogutse, O.~P. 1978, Nuclear Fusion, 18, 303, aDS Bibcode: 1978NucFu..18..303P

\bibitem[{Pierrard \& Lazar(2010)}]{pierrard_kappa_2010}
Pierrard, V. \& Lazar, M. 2010, Solar Physics, 267, 153

\bibitem[{Pierrard {et~al.}(2011)Pierrard, Lazar, \& Schlickeiser}]{pierrard_evolution_2011}
Pierrard, V., Lazar, M., \& Schlickeiser, R. 2011, Solar Physics, 269, 421

\bibitem[{Pierrard {et~al.}(2022)Pierrard, Lazar, \& Stverak}]{pierrard_implications_2022}
Pierrard, V., Lazar, M., \& Stverak, S. 2022, Frontiers in Astronomy and Space Sciences, 9, 892236

\bibitem[{Pierrard {et~al.}(1999)Pierrard, Maksimovic, \& Lemaire}]{pierrard_electron_1999}
Pierrard, V., Maksimovic, M., \& Lemaire, J. 1999, Journal of Geophysical Research: Space Physics, 104, 17021, \_eprint: https://onlinelibrary.wiley.com/doi/pdf/10.1029/1999JA900169

\bibitem[{Pilipp {et~al.}(1987)Pilipp, Miggenrieder, Montgomery, Rosenbauer, \& Schwenn}]{pilipp_characteristics_1987}
Pilipp, W.~G., Miggenrieder, H., Montgomery, M.~D., {et~al.} 1987, Journal of Geophysical Research, 92, 1075

\bibitem[{Pistinner \& Eichler(1998)}]{pistinner_self-inhibiting_1998}
Pistinner, S.~L. \& Eichler, D. 1998, Monthly Notices of the Royal Astronomical Society, 301, 49

\bibitem[{Raouafi {et~al.}(2023)Raouafi, Matteini, Squire, Badman, Velli, Klein, Chen, Matthaeus, Szabo, Linton, Allen, Szalay, Bruno, Decker, Akhavan-Tafti, Agapitov, Bale, Bandyopadhyay, Battams, Bercic, Bourouaine, Bowen, Cattell, Chandran, Chhiber, Cohen, D'Amicis, Giacalone, Hess, Howard, Horbury, Jagarlamudi, Joyce, Kasper, Kinnison, Laker, Liewer, Malaspina, Mann, McComas, Niembro-Hernandez, Panasenco, Pusack, Pulupa, Perez, Riley, Rouillard, Shi, Stenborg, Tenerani, Verniero, Viall, Vourlidas, Wood, Woodham, \& Woolley}]{raouafi_parker_2023}
Raouafi, N.~E., Matteini, L., Squire, J., {et~al.} 2023, Parker {Solar} {Probe}: {Four} {Years} of {Discoveries} at {Solar} {Cycle} {Minimum}, arXiv:2301.02727 [astro-ph, physics:physics]

\bibitem[{Roberg-Clark {et~al.}(2019)Roberg-Clark, Agapitov, Drake, \& Swisdak}]{roberg-clark_scattering_2019}
Roberg-Clark, G.~T., Agapitov, O., Drake, J.~F., \& Swisdak, M. 2019, The Astrophysical Journal, 887, 190

\bibitem[{Roberg-Clark {et~al.}(2018)Roberg-Clark, Drake, Reynolds, \& Swisdak}]{roberg-clark_suppression_2018}
Roberg-Clark, G.~T., Drake, J.~F., Reynolds, C.~S., \& Swisdak, M. 2018, Physical Review Letters, 120, 035101, arXiv:1709.00057 [astro-ph, physics:physics]

\bibitem[{Rosenbauer {et~al.}(1976)Rosenbauer, Miggenrieder, Montgomery, \& Schwenn}]{rosenbauer_preliminary_1976}
Rosenbauer, H., Miggenrieder, H., Montgomery, M., \& Schwenn, R. 1976, in Physics of {Solar} {Planetary} {Environments}: {Proceedings} of the {International} {Symposium} on {Solar}‚{Terrestrial} {Physics}, {June} 7‚18,1976 {Boulder}, {Colorado}, {Volume} {I} (American Geophysical Union (AGU)), 319--331

\bibitem[{Rosenbauer {et~al.}(1977)Rosenbauer, Schwenn, Marsch, Meyer, Miggenrieder, Montgomery, Muehlhaeuser, Pilipp, Voges, \& Zink}]{rosenbauer_survey_1977}
Rosenbauer, H., Schwenn, R., Marsch, E., {et~al.} 1977, Journal of Geophysics Zeitschrift Geophysik, 42, 561

\bibitem[{Sagdeev \& Shafranov(1960)}]{sagdeev_instability_1960}
Sagdeev, R.~Z. \& Shafranov, V.~D. 1960, Zhur. Fiz., Vol: 39

\bibitem[{Saito \& Gary(2007)}]{saito_all_2007}
Saito, S. \& Gary, S.~P. 2007, Geophysical Research Letters, 34, L01102

\bibitem[{Santolik {et~al.}(2003)Santolik, Parrot, \& Lefeuvre}]{santolik_singular_2003}
Santolik, O., Parrot, M., \& Lefeuvre, F. 2003, Radio Science, 38, n/a

\bibitem[{Sarfraz \& Yoon(2020)}]{sarfraz_combined_2020}
Sarfraz, M. \& Yoon, P.~H. 2020, Journal of Geophysical Research: Space Physics, 125

\bibitem[{Schroeder {et~al.}(2021)Schroeder, Boldyrev, \& Astfalk}]{schroeder_stability_2021}
Schroeder, J.~M., Boldyrev, S., \& Astfalk, P. 2021, Monthly Notices of the Royal Astronomical Society, 507, 1329

\bibitem[{Scime {et~al.}(1994)Scime, Bame, Feldman, Gary, Phillips, \& Balogh}]{scime_regulation_1994}
Scime, E.~E., Bame, S.~J., Feldman, W.~C., {et~al.} 1994, Journal of Geophysical Research, 99, 23401

\bibitem[{Scudder(1992{\natexlab{a}})}]{scudder_causes_1992}
Scudder, J.~D. 1992{\natexlab{a}}, The Astrophysical Journal, 398, 299, aDS Bibcode: 1992ApJ...398..299S

\bibitem[{Scudder(1992{\natexlab{b}})}]{scudder_why_1992}
Scudder, J.~D. 1992{\natexlab{b}}, The Astrophysical Journal, 398, 319, aDS Bibcode: 1992ApJ...398..319S

\bibitem[{Shaaban {et~al.}(2018)Shaaban, Lazar, \& Poedts}]{shaaban_clarifying_2018}
Shaaban, S.~M., Lazar, M., \& Poedts, S. 2018, Monthly Notices of the Royal Astronomical Society, 480, 310

\bibitem[{Shaaban {et~al.}(2019)Shaaban, Lazar, Yoon, Poedts, \& Lopez}]{shaaban_quasi-linear_2019}
Shaaban, S.~M., Lazar, M., Yoon, P.~H., Poedts, S., \& Lopez, R.~A. 2019, Monthly Notices of the Royal Astronomical Society, 486, 4498

\bibitem[{Sonnerup \& Cahill~Jr.(1967)}]{sonnerup_magnetopause_1967}
Sonnerup, B. U. \& Cahill~Jr., L.~J. 1967, Journal of Geophysical Research (1896-1977), 72, 171

\bibitem[{Sonnerup \& Scheible(1998)}]{sonnerup_minimum_1998}
Sonnerup, B. U. \& Scheible, M. 1998, ISSI Scientific Reports Series, 1, 185

\bibitem[{Stansby {et~al.}(2016)Stansby, Horbury, Chen, \& Matteini}]{stansby_experimental_2016}
Stansby, D., Horbury, T.~S., Chen, C. H.~K., \& Matteini, L. 2016, The Astrophysical Journal, 829, L16

\bibitem[{Steinvall {et~al.}(2021)Steinvall, Khotyaintsev, Cozzani, Vaivads, Yordanova, Eriksson, Edberg, Maksimovic, Bale, Chust, Krasnoselskikh, Kretzschmar, Lorfevre, Plettemeier, Soucek, Steller, Steverak, Vecchio, Horbury, OBrien, Evans, Fedorov, Louarn, Lavraud, Rouillard, \& Owen}]{steinvall_solar_2021}
Steinvall, K., Khotyaintsev, Y.~V., Cozzani, G., {et~al.} 2021, Astronomy \& Astrophysics, 656, A9, arXiv:2104.03553 [astro-ph, physics:physics]

\bibitem[{Taubenschuss \& Santolik(2019)}]{taubenschuss_wave_2019}
Taubenschuss, U. \& Santolik, O. 2019, Surveys in Geophysics, 40, 39

\bibitem[{Tong {et~al.}(2019{\natexlab{a}})Tong, Vasko, Artemyev, Bale, \& Mozer}]{tong_statistical_2019}
Tong, Y., Vasko, I., Artemyev, A., Bale, S., \& Mozer, F. 2019{\natexlab{a}}, The Astrophysical Journal, 878, 41

\bibitem[{Tong {et~al.}(2019{\natexlab{b}})Tong, Vasko, Pulupa, Mozer, Bale, Artemyev, \& Krasnoselskikh}]{tong_whistler_2019}
Tong, Y., Vasko, I.~Y., Pulupa, M., {et~al.} 2019{\natexlab{b}}, The Astrophysical Journal, 870, L6, aDS Bibcode: 2019ApJ...870L...6T

\bibitem[{Vasko {et~al.}(2019)Vasko, Krasnoselskikh, Tong, Bale, Bonnell, \& Mozer}]{vasko_whistler_2019}
Vasko, I.~Y., Krasnoselskikh, V., Tong, Y., {et~al.} 2019, The Astrophysical Journal, 871, L29

\bibitem[{Vasko {et~al.}(2020)Vasko, Kuzichev, Artemyev, Bale, Bonnell, \& Mozer}]{vasko_quasi-parallel_2020}
Vasko, I.~Y., Kuzichev, I.~V., Artemyev, A.~V., {et~al.} 2020, Physics of Plasmas, 27, 082902

\bibitem[{Vedenov(1963)}]{vedenov_quasi-linear_1963}
Vedenov, A.~A. 1963, Journal of Nuclear Energy. Part C, Plasma Physics, Accelerators, Thermonuclear Research, 5, 169

\bibitem[{Verscharen {et~al.}(2022)Verscharen, Chandran, Boella, Halekas, Innocenti, Jagarlamudi, Micera, Pierrard, Steverak, Vasko, Velli, \& Whittlesey}]{verscharen_electron-driven_2022}
Verscharen, D., Chandran, B. D.~G., Boella, E., {et~al.} 2022, Frontiers in Astronomy and Space Sciences, 9, 951628

\bibitem[{Verscharen {et~al.}(2019)Verscharen, Chandran, Jeong, Salem, Pulupa, \& Bale}]{verscharen_self-induced_2019}
Verscharen, D., Chandran, B. D.~G., Jeong, S.-Y., {et~al.} 2019, The Astrophysical Journal, 886, 136

\bibitem[{Vocks {et~al.}(2005)Vocks, Salem, Lin, \& Mann}]{vocks_electron_2005}
Vocks, C., Salem, C., Lin, R.~P., \& Mann, G. 2005, The Astrophysical Journal, 627, 540

\bibitem[{Yakimenko(1963)}]{yakimenko_absorption_1963}
Yakimenko, V.~L. 1963, Journal of Experimental and Theoretical Physics, 17

\bibitem[{Steverak {et~al.}(2009)Steverak, Maksimovic, Travnicek, Marsch, Fazakerley, \& Scime}]{stverak_radial_2009}
Steverak, S., Maksimovic, M., Travnicek, P.~M., {et~al.} 2009, Journal of Geophysical Research: Space Physics, 114, n/a

\bibitem[{Steverak {et~al.}(2008)Steverak, Travnicek, Maksimovic, Marsch, Fazakerley, \& Scime}]{stverak_electron_2008}
Steverak, S., Travnicek, P., Maksimovic, M., {et~al.} 2008, Journal of Geophysical Research: Space Physics, 113

\bibitem[{Steverak {et~al.}(2015)Steverak, Travnicek, \& Hellinger}]{stverak_electron_2015}
Steverak, S., Travnicek, P.~M., \& Hellinger, P. 2015, Journal of Geophysical Research: Space Physics, 120, 8177

\end{thebibliography}

\begin{appendix}
\section{}
\label{Annexe A}
The X component of the Poynting vector is defined as:
\begin{equation}
    S_{\rm X} = \frac{\biggl(E{\rm w}_{\rm Y}B^{*}{\rm w}_{\rm Z} \biggr)(\omega,t) - \biggl(E{\rm w}_{\rm Z}B^{*}{\rm w}_{\rm Y}\biggr)(\omega,t)}{2 \mu_{0}}
\label{Poyting}
\end{equation}
where $B{\rm w}$ and $E{\rm w}$ are the magnetic and electric waveforms, respectively. $(\omega,t)$ stands for the time average Fourier component and $^{*}$ for the complex conjugate \citep{chust_observations_2021}. $\mu_{0}$ is the vacuum permeability ($\si{\newton} \si{\ampere}^{-2}$). A positive value of $S_{\rm X}$ indicates a direction towards the Sun. 
As explained in \cite{kretzschmar_whistler_2021}, there is an unexpected and nearly constant instrumental phase shift between the electric and magnetic field in the RPW Low Frequency Range (LFR) measurements. In their work, they derived the effective electric antenna length using the whistler wave dispersion relation and compared their results with the value independently obtained by \cite{steinvall_solar_2021} using a deHoffmann-Teller analysis. They showed that $E{\rm w}_{\rm Y}$ (resp., $B{\rm w}_{\rm Z}$) should be delayed (resp., advanced) by removing (resp., adding) $50^\circ$. This phase shift was well established in the case of quasi-aligned whistler waves. A constant $50^\circ$ phase shift correction would not change the sign of $S_{\rm X}$ for a perfect wave having a theoretical phase shift between the magnetic and electric field equal to $0^\circ$ or $180^\circ$ (for sunward and anti-sunward propagation, respectively). But as was shown by \cite{kretzschmar_whistler_2021} the phase shift is not perfectly constant (small dependence with density) and there are of course uncertainties in the measurements. 
We therefore pay special attention to the determination of the direction of propagation with Solar Orbiter data, as described below.\\
First we consider only the cases with $k_{\rm X}/|k| \ge 15\%$ (where $k$ is the wave vector) and with the coherence between $(E{\rm w}_{\rm Y},B{\rm w}_{\rm Z})$ and $(E{\rm w}_{\rm Z},B{\rm w}_{\rm Y})$ greater than 0.6. Then, we look at the phase of $S_{\rm X}$ ($\varphi_{S_{\rm X}}$) (spectral energy content weighted average). With a constant $50^\circ$ phase shift, the observed value of $\varphi_{S_{\rm X}}$ is expected to be $\varphi_{S_X} = 50^\circ$ and  $\varphi_{S_X} = -130^\circ$  for  sunward and anti-sunward propagation, respectively. Taking into account this weak phase deviation, we categorize the cases as sunward propagating if $0^\circ \le \varphi_{S_{\rm X}} \le 100^\circ$ and anti-sunward propagating if $-180^\circ \le \varphi_{S_X} \le -80^\circ$. These restrictive constraints leave a large number of cases unresolved (hereafter named poorly defined) but ensure when they are verified that the direction of propagation is correctly characterized.\\

 
 
 
 
 

\section{}

\label{Annexe B}

To calculate the diffusion coefficients we need: $B{\rm w}$, $\omega_{\rm m}$, $\delta_\omega$, $\omega_{\rm lc}$, $\omega_{\rm uc}$, $X_{\rm m}$,  $X{\rm w}$, $X_{\rm min}$ and $X_{\rm max}$, defined in \cite{glauert_calculation_2005}. These parameters are estimated using the statistics on whistler waves: 
\begin{itemize}
 \item $B{\rm w}$ is the total wave amplitude
 \item $\omega_{\rm m}$ is the frequency at which the signal is maximum
  \item $\omega_{\rm lc}$ is the lowest frequency at which planarity and ellipticity are greater than 0.6
  \item $\omega_{\rm uc}$ is the highest frequency at which planarity and ellipticity are greater than 0.6
   \item $\delta_{\omega} =(w_{\rm uc} - w_{\rm lc})/4$ 
   \item $X_{\rm m}$ is calculated with minimum variance analysis and the spectral matrices (mean value)
   \item $X{\rm w}= tan(3^\circ)$
   \item $X_{\rm min}=X_{\rm m} - tan(5^\circ) $
   \item $X_{\rm max}= X_{\rm m} + tan(5^\circ)$

The values used for ($X{\rm w}$,$X_{\rm min}$,$X_{\rm max}$), that is, 3 and 5$^\circ$ were obtained using typical repartition of the propagation angle with frequency using spectral matrices and power spectral densities. They mainly modify the range of pitch angles that are in resonances and since we use interpolation at the Strahl PAW do not greatly impact our results.

\end{itemize}

\end{appendix}

\end{document}